\documentclass[fleqn,usenatbib]{mnras}

\usepackage{newtxtext,newtxmath}
\usepackage{graphicx}	
\usepackage{amsmath}	
\usepackage{enumerate}
\usepackage{tablefootnote}
\usepackage[dvipsnames]{xcolor} 

\title[Constraining BBH pathways]{Constraints on the contributions to the observed binary black hole population from individual evolutionary pathways in isolated binary evolution}

\author[Stevenson \& Clarke]{
\newauthor  Simon Stevenson$^{1,2}$, 
and Teagan A. Clarke$^{2,3}$\\
$^{1}$ Centre for Astrophysics and Supercomputing, Swinburne University of Technology, John St, Hawthorn, Victoria- 3122, Australia \\
$^{2}$ The ARC Centre of Excellence for Gravitational Wave Discovery,  OzGrav \\
$^{3}$ School of Physics and Astronomy, Monash University, Clayton, Vic. 3800, Australia
}

\date{Accepted 2022 October 10. Received 2022 October 9; in original form 2022 August 10}

\pubyear{2022}

\begin{document}
\label{firstpage}
\pagerange{\pageref{firstpage}--\pageref{lastpage}}
\maketitle

\begin{abstract}
Gravitational waves from merging binary black holes can be used to shed light on poorly understood aspects of massive binary stellar evolution, such as the evolution of massive stars (including their mass-loss rates), the common envelope phase, and the rate at which massive stars form throughout the cosmic history of the Universe.
In this paper we explore the \emph{correlated} impact of these phases on predictions for the merger rate and chirp mass distribution of merging binary black holes, aiming to identify possible degeneracies between model parameters.
In many of our models, a large fraction (more than $70$\% of detectable binary black holes) arise from the chemically homogeneous evolution scenario; these models tend to over-predict the binary black hole merger rate and produce systems which are on average too massive.
Our preferred models favour enhanced mass-loss rates for helium rich Wolf--Rayet stars, in tension with recent theoretical and observational developments.
We identify correlations between the impact of the  mass-loss rates of Wolf--Rayet stars and the metallicity evolution of the Universe on the rates and properties of merging binary black holes.
Based on the observed mass distribution, we argue that the $\sim 10\%$ of binary black holes with chirp masses greater than $40$\,M$_\odot$ (the maximum predicted by our models) are unlikely to have formed through isolated binary evolution, implying a significant contribution ($> 10$\%) from other formation channels such as dense star clusters or active galactic nuclei.
Our models will enable inference on the uncertain parameters governing binary evolution in the near future.
\end{abstract}




\section{Introduction}
\label{sec:intro}

Almost 100 binary black hole mergers have now been observed by the LIGO
\citep{TheLIGOScientificDetector:2014jea} and Virgo \citep{TheVirgoDetector:2014hva} gravitational-wave observatories, as identified and published in a series of catalogues by the LIGO, Virgo and KAGRA Scientic Collaborations \citep{LIGOScientific:2018mvr,Abbott:2021PhRvXGWTC-2,LIGOScientific:2021usb,LIGOScientific:2021djp}. 
A number of additional candidates have been identified in analyses of the data by external groups \citep[e.g.,][]{Nitz:2021zwj,Olsen:2022pin}.

Analyses of the compact binary population using phenomenological population models have allowed the merger rates, mass and spin distributions of binary black holes to be constrained \citep{LIGOScientific:2021psn}.
For example, \citet{LIGOScientific:2021psn} find that the rate of binary black hole mergers at redshift $z = 0.2$ is 17--45\,Gpc$^{-3}$\,yr$^{-1}$, and increases with redshift, in agreement with the measured cosmic star formation history \citep{Madau:2014bja}.
However, beyond identifying some large scale properties of the population, it is difficult to relate these features to the underlying physics describing the formation of binary black holes.

The origin of these gravitational-wave sources is currently a major open question in astrophysics. 
Several potential formation scenarios have been proposed as being able to explain the rates and properties of the observed binary black hole mergers (for a recent high-level overview of binary black hole formation scenarios, see \citealp{Mandel:2018hfr}).
These scenarios include the evolution of massive binary stars \citep{Belczynski:2016obo,Stevenson:2017tfq}, and the dynamical formation of black hole binaries in star clusters \citep[e.g.,][]{Rodriguez:2016kxx,DiCarlo:2020lfa} or active galactic nuclei \citep[e.g.,][]{Yang:2020lhq}. 
These scenarios make different predictions for the properties of merging binary black holes, and predict a wide range of merger rates \citep{Mandel:2022LRR}.

In this paper, we focus our attention on modelling the formation of binary black holes through isolated binary evolution \citep{Belczynski:2016obo,Stevenson:2017tfq}. 
However, even within this channel, there are a number of different possible evolutionary pathways which may contribute to the observed binary black hole population.
The most well known scenario involves the evolution of initially wide binaries, that interact through phases of mass transfer and/or common envelope evolution that shrink the orbit of the binary, finally forming a binary black hole that can merge due to the emission of gravitational waves within the age of the Universe. 

Another evolutionary pathway that has been explored recently is the formation of merging binary black holes through chemically homogeneous evolution \citep[][]{Mandel:2015qlu,Marchant:2016wow,duBuisson:2020asn,Riley:2020btf}.
In this channel, stars born in binaries with very short orbital periods can become tidally locked, leading to them rotating rapidly, enhancing mixing within the star.
This can lead to chemically homogeneous evolution \citep{Maeder:1987A&A}, where essentially the entire star is converted to helium through nuclear burning.
This avoids the phases of radial expansion experienced by more conventional stellar evolution pathways, and may potentially allow for the formation of massive binary black holes in low-metallicity environments.
Each of these channels produces sub-populations of binary black holes with different characteristics (such as masses).

There are many uncertainties in the evolution of massive binary stars \citep[e.g.,][]{Belczynski:2022ApJ}, such as which binaries enter and survive the common envelope phase (and what their final orbital properties are), what the final masses and kicks of compact objects formed in supernovae are, and what the mass loss rates of evolved, low metallicity massive stars are, among others.
We discuss each of these phases of binary evolution in greater detail in Section~\ref{sec:methods}.

The impact of these uncertainties on the predictions for binary black hole formation have been extensively studied in the literature, typically through the use of binary population synthesis models.
Population synthesis codes commonly used to model gravitational-wave sources include \textsc{StarTrack} \citep{Belczynski:2016obo}, \textsc{Mobse} \citep[][]{Giacobbo:2017qhh} and \textsc{Cosmic} \citep[][]{Breivik:2019lmt}.
Many works have focused on varying a single parameter (corresponding to a single uncertain stage of binary evolution) at a time \citep[e.g.,][]{Voss:2003ep,Dominik:2012kk,Kruckow:2018slo}.
This strategy has been successful as it has a number of positive aspects.
Restricting variations to one piece of physics at a time keeps the computational cost of these analyses under control. 
In addition, isolating the impact of a single piece of physics on binary evolution aids tremendously in interpreting the results of these controlled numerical experiments.
The predictions from these population synthesis models can then be compared to gravitational-wave observations in order to place constraints on the underlying binary evolution processes \citep[][]{Stevenson:2015bqa}. 

For example, following the first observations of neutron star-black hole mergers \citep{LIGOScientific:2021qlt}, \citet{Broekgaarden:2021hlu} compared the observed rates of binary neutron star, neutron star-black hole and binary black hole coalescences \citep{LIGOScientific:2021psn} to a suite of predictions from the set of population synthesis models from \citet{Broekgaarden:2021iew,Broekgaarden:2021efa}. 
They find that only models with low supernova kicks or high common envelope efficiencies match the observed compact binary merger rates.
Similar conclusions were reached by \citet{Santoliquido:2020axb}.

\begin{figure}
    \centering
    \includegraphics[width=\columnwidth]{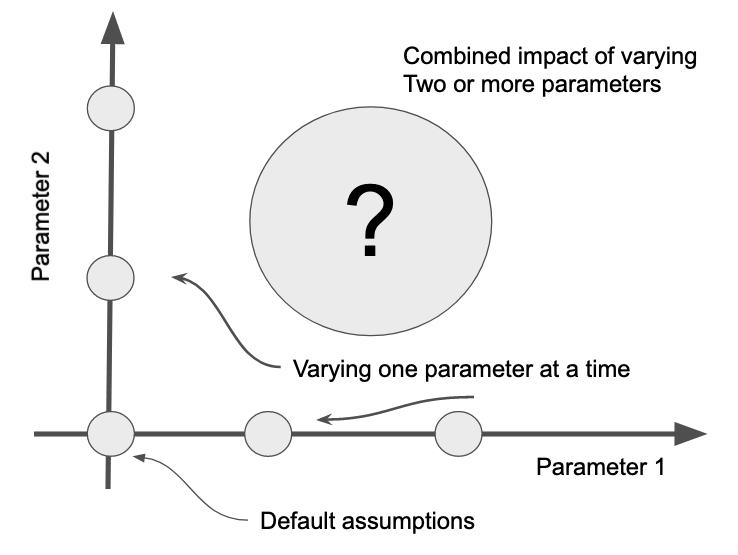}
    \caption{Cartoon illustrating the common method of exploring the binary evolution parameter space by varying a single parameter at a time. 
    The gray circles represent the locations in the parameter space that models are computed. 
    The large gray circle with a question mark highlights the parameter space unexplored with that method. 
    Exploring this parameter space is the main aim of this work.}
    \label{fig:parameter_space_cartoon}
\end{figure}

However, this method of varying one parameter (or one assumption) at a time is clearly not sufficient to fully explore the model parameter space (see Figure~\ref{fig:parameter_space_cartoon}). 
Any conclusions reached by comparing observations to a limited set of models may therefore be biased. 
This is because the combined impact of varying several different parameters simultaneously has not been explored, and two (or more) parameters may counteract one another, resulting in a \emph{degeneracy} between multiple model parameters, such that the model results may be unchanged.
Exploring these possibilities is the main aim of the present work.

\citet{Barrett:2017fcw} explored correlations between the impact of several binary evolution processes on the population of binary black holes predicted from isolated binary evolution. 
In particular, they used a Fisher matrix analysis to examine the correlations between the impact of the efficiency of the common envelope phase, supernova kicks and wind mass loss rates on the population of binary black holes. 
\citet{Barrett:2017fcw} explored how the predictions of their model varied under variations of four population parameters by a few percent around their values in a fiducial model using a method similar to that illustrated in Figure~\ref{fig:parameter_space_cartoon}.
They demonstrated that $\sim 1000$ gravitational-wave observations---a target expected to be reached in the next few years---would be sufficient to allow for precise measurements of many of these parameters.
They found several interesting correlations, such as between the efficiency of common envelope evolution and the mass loss rates of Wolf--Rayet stars (see Section~\ref{sec:methods} for more details of the parameterisation of these phases).
Whilst \citet{Barrett:2017fcw} demonstrated the viability of precisely constraining population parameters using gravitational-wave observations, and highlighted some interesting correlations between population parameters.
However, the analysis of \citet{Barrett:2017fcw} was restricted to a small region of parameter space localised around a fiducial model, and thus does not give a clear picture of how the predictions of the model would vary across the broader parameter space.
In addition, they assumed a single model for the cosmic star formation history of the universe, and did not account for the uncertainties there, which have since been shown to have an important impact on predictions for gravitational-wave mergers \citep[e.g.,][]{Neijssel:2019,Broekgaarden:2021efa}.
Finally, \citet{Barrett:2017fcw} did not self consistently include the chemically homogeneous evolution pathway for forming binary black holes in their modelling.

With respect to uncertainties in the cosmic star formation history, \citet{Broekgaarden:2021iew,Broekgaarden:2021efa} took a pioneering step by examining the impact of uncertainties in both the cosmic star formation rate history and massive binary evolution simultaneously (see also \citealt{Santoliquido:2020axb}). 
However they still only varied one binary evolution parameter at a time (with one exception of one model).

In this paper we build upon these earlier works and examine the correlated impact of uncertainties in both massive binary stellar evolution and the cosmic star formation history on the properties of binary black hole mergers observable in gravitational waves. 
Understanding correlations in the space of models should ultimately aid in interpreting the constraints obtained on population synthesis models through comparison with observations \citep[e.g.,][]{Bouffanais:2020qds}.

The remainder of this paper is structured as follows: in Section~\ref{sec:methods} we describe the population synthesis code that we use to simulate populations of merging binary black holes, and give details of the parameterisations we employ for common envelope evolution, the mass-loss rates of Wolf--Rayet stars and the cosmic star formation history of the Universe, as well as of our model exploration.
In Section~\ref{sec:observational_sample} we summarise the currently observed sample of binary black hole mergers.
We argue that the observed chirp mass distribution already appears to show evidence of contributions from multiple formation scenarios, as $\sim 10\%$ of binary black hole mergers are too massive to have formed through isolated binary evolution in our model, and must have formed through alternate channels such as in dense star clusters \citep[e.g.,][]{Rodriguez:2017pec} or active galactic nuclei \citep[e.g.,][]{Yang:2019cbr}. 
We describe the predictions of our large suite of $2,916$ models in Section~\ref{sec:results}, focusing on demonstrating the range of binary black hole merger rates (Section~\ref{subsec:BBH_rates}) and mass distributions (Section~\ref{subsec:BBH_mass_dist}) that are possible to attain through isolated binary evolution alone.
We place particular emphasis on the contribution from binaries formed through chemically homogeneous evolution to the total population.
We compare our models to current gravitational-wave observations in Section~\ref{subsec:compare_to_obs}.
We find that our preferred models have enhanced mass-loss rates for massive, evolved, helium-rich Wolf--Rayet stars.
We also show that none of our models can produce a satisfactory match to the observed chirp mass distribution, and discuss some possible reasons for this discrepancy.
Finally, we conclude and discuss limitations of the present work, and avenues for future work in Section~\ref{sec:conclusions}.

\section{Method}
\label{sec:methods}

Most massive stars are known to be in binary or higher order multiple systems \citep{Sana:2012Science,MoeDiStefano:2017ApJS}.
We model the evolution of massive isolated binaries using the rapid binary population synthesis suite COMPAS \citep{Stevenson:2017tfq,Vigna-Gomez:2018dza,COMPAS:2021methodsPaper}.

The evolution of massive stars in COMPAS is approximated using the Single Stellar Evolution \citep[SSE;][]{Hurley:2000MNRASSSE} package, a set of analytic polynomial expressions fit to the set of detailed stellar models from \citet{Pols:1998MNRAS}. The SSE package provides a fast and robust method to estimate parameters of stars such as their radii, luminosities and evolutionary timescales.

The implementation of binary evolution (e.g., mass transfer, common envelope evolution) in COMPAS broadly follows the prescriptions described in \citet{Hurley:2002MNRASBSE}, with modifications as described in \citet{COMPAS:2021methodsPaper}.
With the exception of the special case of chemically homogeneous evolution \citep{Riley:2020btf}, tides are not currently implemented in COMPAS \citep{COMPAS:2021methodsPaper}.
We briefly summarise the implementation of chemically homogeneous evolution in COMPAS in Section~\ref{subsec:CHE} below.

The initial parameters of our binaries are drawn from simple distributions, motivated by observations of massive binary stars.
We draw the initial mass of the more massive star in the binary (the primary) from an initial mass function \citep{Kroupa:2000iv} with a power-law slope of $-2.3$ for masses between 1\,M$_\odot$ and $150$\,M$_\odot$.
We draw the initial separations of the binaries from a log-uniform distribution between 0.01 and 1000\,AU\footnote{\citet{Sana:2012Science} find a separation distribution which favours close binaries compared to our default separation distribution, which may be favourable for chemically homogeneous evolution. In this sense, our results for the fraction of binary black holes formed through chemically homogeneous evolution in Section~\ref{sec:results} may be considered lower limits.}.
The mass ratio of the binary is then drawn from a uniform distribution \citep{Sana:2012Science}, with the restriction that the mass of the secondary star must be greater than 0.1\,M$_\odot$.
In order to reduce the initial parameter space, we assume that all massive binaries are initially circular; whilst this is not representative of the eccentricity distribution of massive binaries in nature, studies have shown that the impact on the results from binary population synthesis simulations is mild \citep{Hurley:2002MNRASBSE,deMink:2015yea}. 
We have also assumed that the distribution of initial binary properties is separable, which may not be the case \citep{MoeDiStefano:2017ApJS}, although studies have shown that properly accounting for this does not dramatically affect the predictions from population synthesis models \citep{Klencki:2018zrz}.

As discussed in the introduction, there are several different evolutionary pathways for binary black hole formation within the isolated binary evolution channel.
Our COMPAS population synthesis models include the formation of merging binary black holes through stable mass transfer after the main sequence, common envelope evolution and chemically homogeneous evolution. 
However, it is worth nothing that there may still be additional channels that contribute to the formation of binary black holes that are not currently modelled within COMPAS, such as formation through stable mass transfer on the main sequence (so called case A mass transfer) \citep[][]{Valsecchi:2010Natur,Qin:2018sxk,Neijssel:2021imj}, though we do not expect this to be the dominant channel \citep[cf.][]{Gallegos-Garcia:2022arXiv}.

We use the remnant prescription from \citet{Fryer:2012ApJ} to determine the masses of black holes from the properties of a star (such as its total mass and core mass) at the time of core-collapse. 
In this model, most heavy black holes form through complete fallback with no associated kick. 
We assume that 0.1\,M$_\odot$ is lost through neutrinos in the collapse to a black hole \citep{Stevenson:2019rcw}.

Stellar evolution calculations make a robust prediction of a gap in the mass spectrum of black holes due to the effects of pair-instability supernovae \citep{Belczynski:2016jno,Marchant:2019ApJ,Stevenson:2019rcw,Farmer:2019ApJ}.
The exact location of this gap depends on the details of the stellar models, but typically models predict an absence of black holes in the mass range 50--120\,M$_\odot$ \citep[e.g.,][]{Woosley:2016hmi}, whilst in some cases, stars that retain a massive hydrogen envelope may potentially produce black holes up to and beyond 70\,M$_\odot$ \citep[][]{Costa:2020MNRAS,Renzo:2020smh,Farrell:2020zju}. 
Such stars are unlikely to exist in the compact binaries we consider here, as their hydrogen envelopes are likely removed through binary interactions.
The effects of pair-instability supernovae are implemented in COMPAS as detailed in \citet{Stevenson:2019rcw}, using a fit to the detailed models of \citet{Marchant:2019ApJ}.
In our models, we find that the most massive black hole below the pair-instability mass gap has a mass of around $45$\,M$_\odot$.
We choose to focus our investigation on binary black holes where both black holes have masses below the mass gap for a number of reasons:
1) Only a few very massive stars ($> 150$\,M$_\odot$) are known \citep[e.g.,][]{Figer:2005Natur,Crowther:2010MNRAS,Bestenlehner:2011A&A},  their statistics are highly uncertain, and they may be the result of stellar mergers \citep[e.g.,][]{Banerjee:2012MNRAS} 
2) Stellar models for very massive stars are highly uncertain \citep[e.g.,][]{Agrawal:2021arXiv} 
3) There are currently no robust observations of gravitational-wave sources from above the mass-gap \citep[][]{LIGOScientific:2021djp}.

As discussed above, our current understanding of several stages of massive binary evolution remain uncertain. 
In the following subsections we describe the prescriptions that we adopt for several of these phases, namely common envelope evolution, the mass-loss rates of massive stars, and the cosmic star formation history of the Universe.
We give details of our default assumptions, as well as the variations that we consider.

\subsection{Chemically homogeneous evolution}
\label{subsec:CHE}

Massive close binaries may experience enhanced mixing due to tidal locking, leading to chemically homogeneous evolution. 
In this section we briefly summarise the implementation of chemically homogeneous evolution in COMPAS. For more details, see \citet{Riley:2020btf}.

The angular frequency of a tidally locked star is equal to the orbital frequency. 
It has previously been shown that there is a threshold angular frequency, beyond which stars experience chemically homogeneous evolution.
In COMPAS, this threshold is determined by fits to one-dimensional rotating massive star models, calculated using Modules for Experiments in Stellar Astrophysics \cite[MESA;][]{Paxton:2011ApJS}, as described in \citet{Riley:2020btf}, as a function of mass and metallicity.
If the angular frequency of a star on the zero-age main-sequence exceeds this threshold, it is assumed to evolve chemically homogeneously.
We employ the variant of chemically homogeneous evolution within COMPAS that requires this threshold to be met throughout the main-sequence.

In COMPAS, we assume that the radius of a chemically homogeneously evolving star remains fixed to its zero-age main-sequence radius, whilst its luminosity is assumed to be the same as the luminosity of a normal, slowly rotating main-sequence star of the same fractional age \citep[][]{Hurley:2000MNRASSSE}.

At the end of the main sequence, the star is assumed to evolve directly to the helium main-sequence \citep[as given by][]{Hurley:2000MNRASSSE}, with the same mass as the total mass of the star at the end of the hydrogen main sequence.

\subsection{Common envelope evolution}
\label{subsec:common_envelope}

One of the most famous (and least understood) phases in the evolution of massive binaries is the common envelope phase \citep{Paczynski:1976IAUS,Ivanova:2013A&ARv}, in which unstable mass transfer from an evolved giant star onto a compact star (such as a black hole) leads to the black hole orbiting inside of the envelope of the giant star. 
The black hole spirals in towards the core of the giant star due to drag forces, and in doing so injects energy into the envelope of the giant. 
If sufficient energy is available, the inspiral may be halted (after a dramatic reduction of the binary orbital separation) and the envelope may be ejected.

We make the standard assumption that some fraction $\alpha_\mathrm{CE}$ of the orbital energy can be used to unbind the envelope \citep{Webbink:1984ApJ,deKool:1990ApJ}
\begin{equation}
    E_\mathrm{bind} = \alpha_\mathrm{CE} \Delta E_\mathrm{orb} ,
    \label{eq:alpha_ce}
\end{equation}
where the envelope binding energy is given by
\begin{equation}
    E_\mathrm{bind} = - \frac{G M M_\mathrm{env}}{\lambda R} .
    \label{eq:envelope_binding_energy}
\end{equation}
In this expression, $M$ is the total mass of a star, whilst $M_\mathrm{env}$ is its envelope mass and $R$ its radius. 
The envelope binding energy, $E_\mathrm{bind}$, is parameterised by the dimensionless parameter $\lambda$. 
We use the fitting formulae for $\lambda$, constructed using a set of detailed stellar models, from \citet{XuLi:2010ApJ}.

For the typical common envelope phase of interest en route to the formation of a binary black hole, $\Delta E_\mathrm{orb}$ is given by
\begin{equation}
    \Delta E_\mathrm{orb} = - \frac{G M_\mathrm{BH,pre} M_\mathrm{comp,pre}}{2 a_\mathrm{pre-CE}} - \frac{G M_\mathrm{BH,post} M_\mathrm{comp,post}}{2 a_\mathrm{post-CE}},
    \label{eq:DeltaEorb} 
\end{equation}
where $a_\mathrm{pre-CE}$, $M_\mathrm{BH,pre}$ and $M_\mathrm{comp,pre}$ are the orbital separation, black hole mass and companion mass prior to the common envelope event, and $a_\mathrm{post-CE}$, $M_\mathrm{BH,post}$ and $M_\mathrm{comp,post}$ are the same quantities afterwards.
Larger values of $\alpha_\mathrm{CE}$ represent a more efficient use of the orbital energy to unbind the envelope, resulting in wider post-common envelope binary separations.
Our default model assumes $\alpha_\mathrm{CE} = 1$ for all common envelope events (we discuss this potentially problematic assumption in Section~\ref{sec:conclusions}).

If (as formulated here) the only available source of energy is the orbital energy then $0 \leq \alpha_\mathrm{CE} \leq 1$.
However, there may be additional energy sources available beyond the orbital energy, such as recombination energy \citep{Ivanova:2014tpa,Ivanova:2018wgl,Kruckow:2016A&A,Reichardt:2020MNRAS,Lau:2021jpm} or accretion-powered jets \citep[e.g.,][]{Soker:2018msh,Grichener:2021xeg}, implying that $\alpha_\mathrm{CE} > 1$ may be possible. 
Indeed, some observations of post-common envelope binaries (though these are typically of much lower mass than the systems of interest here) seem to suggest the need for additional energy sources \citep[e.g.,][]{DeMarco:2011MNRAS,Iaconi:2019MNRAS}.
Recent simulations of the common envelope phase in massive binaries have found a range of effective values for $\alpha_\mathrm{CE}$ \citep[][]{Fragos:2019box,Law-Smith:2020jwf,Lau:2021jpm,Moreno:2021otq}.
We therefore explore the range $0.1 \leq \alpha_\mathrm{CE} \leq 10$.
Since we are uncertain a priori whether $\alpha_\mathrm{CE} > 1$ or $\alpha_\mathrm{CE} < 1$, we explore both regimes with a similar number of models.

In addition to the uncertainty of the $\alpha_\mathrm{CE}$ parameter (or equivalently, the mapping between initial conditions and final outcomes of the common envelope phase), there are several other uncertainties associated with common envelope evolution that we do not consider in detail here.
Firstly, estimates of envelope binding energies from detailed stellar models are sensitive to assumptions made in those models, such as the boundary between the core and the envelope of the star \citep{Tauris:2001cx,Kruckow:2016A&A}.
Since the product of $\alpha_\mathrm{CE}$ and $\lambda$ is relevant in determining the outcome of common envelope evolution, our models with different values of $\alpha_\mathrm{CE}$ could also be viewed as probing systematic over/underestimations of the envelope binding energies.
Secondly, modern detailed stellar models of massive stars seem to indicate that ejection of the envelopes of massive, low metallicity stars may only be possible for supergiants in a narrow range of orbital periods 
\citep{Klencki:2020kxd,Marchant:2021hiv}.

An additional uncertainty in common envelope evolution regards the fate of donor stars crossing the Hertzsprung gap.
Stars just beyond the main sequence may not have developed sufficient core-envelope separation to be able to survive the common envelope phase.
Here we assume that all common envelope events involving a donor star on the Hertzsprung gap result in stellar mergers, following \citet{Belczynski:2006zi} and \citet{Dominik:2012kk}.

Since binary black holes formed through the chemically homogeneous evolution pathway do not experience the common envelope phase, we do not expect our treatment of this phase to affect our predictions for this scenario.

\subsection{Massive star winds}
\label{subsec:winds}

Massive stars can lose substantial amounts of mass throughout their lives through strong stellar winds \citep[see][for a review]{Vink:2021arXiv}. 
There are several stages of massive stellar evolution where mass loss can be important, including line-driven winds from hot OB stars on the main sequence \citep{Vink:2001cg}, dust driven winds from cool red supergiants \citep[][]{Mauron:2011A&A,Beasor:2020MNRAS}, and eruptive mass loss from luminous blue variables \citep{Vink:2002A&A,Smith:2017RSPTA}\footnote{Stars evolving chemically homogeneously do not experience the red supergiant or luminous blue variable phases, and thus are not impacted by mass loss during these phases}.
The amount of mass a star loses throughout its life can impact the evolution of massive stars and the final properties of their remnants \citep{Renzo:2017A&A}.

Mass loss rates for stars at each of these different evolutionary stages can be estimated either from observations, or from theory.
For massive, low metallicity stars (the same stars expected to produce many of the observed binary black holes), the mass loss rates are poorly constrained from observation, as massive stars are inherently rare and short lived, and there are only a few local low metallicity environments (such as the Magellanic clouds) where low metallicity stars can be studied in detail. 

We follow the default mass loss prescription in COMPAS \citep{COMPAS:2021methodsPaper} (see also \citealp{Belczynski:2010ApJ}). 
We give more details about the mass loss rates assumed for Wolf--Rayet stars in the section below.

\subsubsection{Wolf--Rayet stars}
\label{subsubsec:WR_mass_loss}

Wolf--Rayet stars\footnote{We do not distinguish between the various subclasses of Wolf--Rayet stars (e.g., WN, WC, WO).} are massive, evolved, helium rich stars with high mass loss rates of $>10^{-5}$\,M$_\odot$\,yr$^{-1}$ \citep[e.g.,][]{Barlow:1981MNRAS}.
They may be formed in single stellar evolution by the removal of the hydrogen envelope of a star through stellar winds, or, perhaps more commonly, can be formed in binary evolution when the hydrogen envelope is removed through mass transfer \citep{Paczynski:1967AcA}.

Mass loss during the Wolf--Rayet stage is important for determining the masses of black holes \citep[][]{Belczynski:2010ApJ,Higgins:2021jux,Vink:2020nak}.
It is particularly important for binary black holes formed through the chemically homogeneous evolution channel \citep{Mandel:2015qlu,Riley:2020btf}.
Firstly, mass loss may lead to a widening of the binary orbit, resulting in binaries no longer being in close enough orbits to maintain high rotation rates, and secondly that mass loss carries away angular momentum, which again may slow the rotation of stars, reducing the parameter space for chemically homogeneous evolution.

Following \citet{Belczynski:2010ApJ} we assume that the mass loss rates of Wolf-Rayet stars are given by
\begin{equation}
    \dot{M}_\mathrm{WR} = f_\mathrm{WR} \times 10^{-13} \left( \frac{L}{L_\odot} \right)^{1.5} \left( \frac{Z}{Z_\odot} \right)^{m} \, \mathrm{M}_\odot \, \mathrm{yr}^{-1} ,
    \label{eq:WR_mass_loss}
\end{equation}
where $L$ is the stellar luminosity, $Z$ is the metallicity and $f_\mathrm{WR}$ is a scaling parameter introduced to allow us to easily vary the mass loss rates during this phase of stellar evolution \citep{Barrett:2017fcw,COMPAS:2021methodsPaper}.
This prescription is derived from the results of theoretical models. 
The dependence of the mass loss rate with luminosity comes from \citet{Hamann:1998A&A}, while the parameter $m = 0.86$ determines the scaling of the mass loss rates with metallicity, and is determined from theoretical models by \citet{Vink:2005zf}. 
Our default choice is $f_\mathrm{WR} = 1$.

Recently, several authors have suggested that Equation~\ref{eq:WR_mass_loss} may either under- or over-estimate the mass loss rates of Wolf--Rayet stars \citep[see e.g., discussion in][in particular their Figure 1]{Sander:2020MNRAS}.

On the theoretical side, \citet{Vink:2017A&A} presented a new set of mass-loss rates for stripped stars that are significantly lower than those found by \citet{Hamann:1998A&A}.
Building on the work of \citet{Vink:2017A&A}, \citet{Sander:2020MNRAS} found that helium star mass loss rates may decrease dramatically below a transition luminosity.

On the empirical side, a popular formula for Wolf--Rayet star mass loss, utilised in many stellar evolution codes, was devised by \citet{Nugis:2000A&A}. 
This expression predicts higher mass loss rates than those from \citet{Hamann:1998A&A} or \citet{Vink:2017A&A}.
Recently, \citet{Tramper:2016ApJ} found even higher mass loss rates than those found by \citet{Nugis:2000A&A}.
\citet{Hamann:2019A&A} utilised improved constraints on the distances to Galactic WN stars from \emph{Gaia} to update their measurements of the mass loss rates. 
They found a large scatter in the mass loss rates as a function of luminosity, with a weaker dependence on luminosity than found by \citet{Nugis:2000A&A}.
\citet{Yoon:2017dme} found that a mild increase in the mass-loss rates of Wolf--Rayet stars by about 60\,\% (i.e., $f_\mathrm{WR} \sim 1.6$) compared to commonly used mass-loss prescriptions improved the agreement between their stellar models and observations of faint WC/WO stars.
\citet{Neijssel:2021imj} argue for reduced wind mass loss rates for helium stars (compared to Equation~\ref{eq:WR_mass_loss}) in order to explain the high black hole mass inferred in the Galactic X-ray binary Cygnus X-1 \citep{Miller-Jones:2021plh}.

\begin{figure}
    \centering
    \includegraphics[width=\columnwidth]{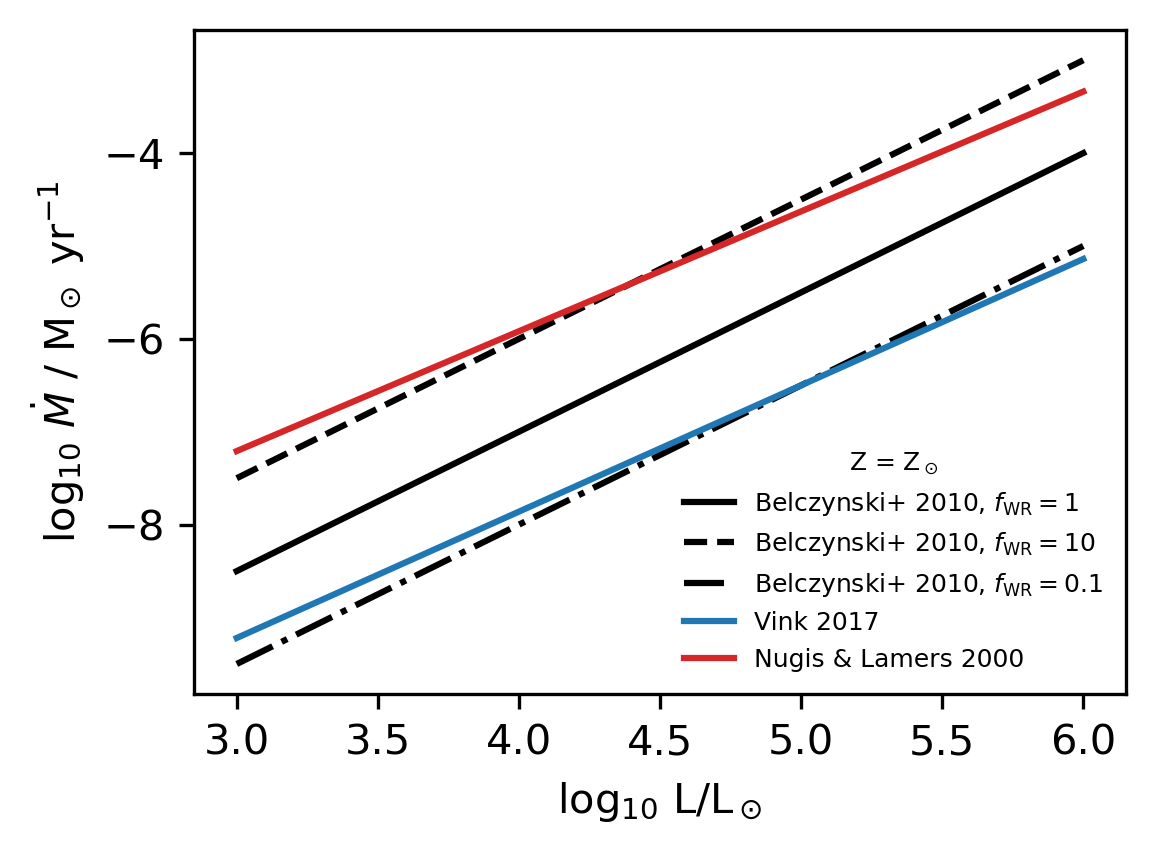} 
    \caption{
    Mass loss rates for WR stars as a function of stellar luminosity at solar metallicity ($Z = Z_\odot$), as predicted by the prescriptions discussed in Section~\ref{subsubsec:WR_mass_loss}. 
    The solid black line denotes our standard assumption for WR mass loss rates ($f_\mathrm{WR} = 1$) \citep{Hamann:1998A&A,Belczynski:2010ApJ}
    The dashed black line shows the same prescription, but with $f_\mathrm{WR} = 10$, whilst the dot-dashed black line shows the values for $f_\mathrm{WR} = 0.1$.
    The solid blue line shows the predictions from \citet{Vink:2017A&A}, whilst the solid red line shows the predictions from \citet{Nugis:2000A&A}.
    }
    \label{fig:WR_mass_loss_rates_func_L}
\end{figure}

We compare the mass-loss rates predicted by these prescriptions as a function of stellar luminosity at solar metallicity ($Z = Z_\odot$) in Figure~\ref{fig:WR_mass_loss_rates_func_L}.
We find that varying $f_\mathrm{WR}$ in the range 0.1--10 sufficiently captures the range of predictions from these models.
Similarly to the case of the common envelope efficiency parameter $\alpha_\mathrm{CE}$, we choose to design our grid of simulations such that we have an equal number of models with $f_\mathrm{WR} > 1$ and $f_\mathrm{WR} < 1$.

\subsection{Cosmic star formation history}
\label{subsec:star_formation_history}

Even though massive stars live for only a few million years, merging compact neutron star and black hole binaries can have long delay times of up to billions of years (Gyr) between the formation of the binary and the eventual gravitational-wave driven merger \citep[][]{Neijssel:2019,Broekgaarden:2021iew}.
This means that compact binaries that merge in the local universe and are observable by LIGO/Virgo may be the products of stars formed at (much) higher redshifts  \citep[e.g.,][]{Belczynski:2016obo,Neijssel:2019}.

The overall star formation rate density budget of the universe is reasonably well constrained at redshifts $z \lesssim 2$, where the star formation rate is observed to increase by a factor of $\sim 10$ between the local universe ($z = 0$) and the peak of star formation around redshift $z = 2$. 
At higher redshifts, the total star formation rate is more poorly constrained observationally, largely due to the impact of dust extinction \citep[][]{Madau:2014bja,Finkelstein:2016PASA}.

One can incorporate the star formation history in a number of ways.
A common method is to make use of analytic or phenomenological models for the cosmic star formation rate density as a function of redshift, calibrated to observations \citep{Madau:2014bja}.
Alternatively, one can attempt to use predictions of the star formation rate at high redshift from cosmological simulations. 
Several groups have pursued this approach, combining the outputs of various combinations of population synthesis models and cosmological simulations to study the population of gravitational-wave mergers \citep[e.g.,][]{Mapelli:2017hqk,O'Shaughnessy:2017MNRAS,Lamberts:2018MNRAS,Mapelli:2018wys,vanSon:2021zpk,Briel:2021bpb}.

In this paper, we opt to use simple phenomenological model of the cosmic star formation rate, which can then be calibrated to match observations. In particular, we use the following description of the cosmic star formation rate density following \citet{Madau:2014bja}

\begin{equation}
    \psi(z) = 
    \frac{{\rm d}^2 M_\mathrm{SFR}}{{\rm d} t_s {\rm d} V{_c}} (z) =  a \frac{(1+z)^b}{1+ [(1+z)/c]^d} \rm M_{\odot} yr^{-1} Mpc^{-3} , 
	\label{eq:SFR}
\end{equation}
where $a$, $b$, $c$ and $d$ are fitting parameters. 
\citet{Madau:2014bja} find that $a = 0.015$, $b = 2.7$, $c = 2.9$ and $d = 5.6$.
\citet{Madau:2016jbv} updated this fit and found $a = 0.01$, $b = 2.6$, $c = 3.2$ and $d = 6.2$.
\citet{Neijssel:2019} calibrated the parameters for Equation~\ref{eq:SFR} by comparing COMPAS models to the rate and mass distribution of binary black holes observed during the first and second observing runs of Advanced LIGO \citep{LIGOScientific:2018mvr}.
They found $a = 0.01$, $b = 2.77$, $c = 2.9$ and $d = 4.7$.
We note however that at that time, COMPAS did not include the chemically homogeneous evolution channel. 
Since the inclusion of this formation channel changes the predicted binary black hole mass distribution and rate \citep{Riley:2020btf}, we expect that the best fit parameters for Equation~\ref{eq:SFR} will change.
At high redshift, the total star formation rate remains highly uncertain, primarily due to the impact of dust extinction.
\citet{Strolger:2004ApJ} found a star formation rate which remains much higher than that given by \citet{Madau:2014bja} at high redshift ($z > 2$).
A similar result was found recently by \citet{Enia:2022arXiv} who studied galaxies using radio observations.

\begin{figure}
    \centering
    \includegraphics[width=\columnwidth]{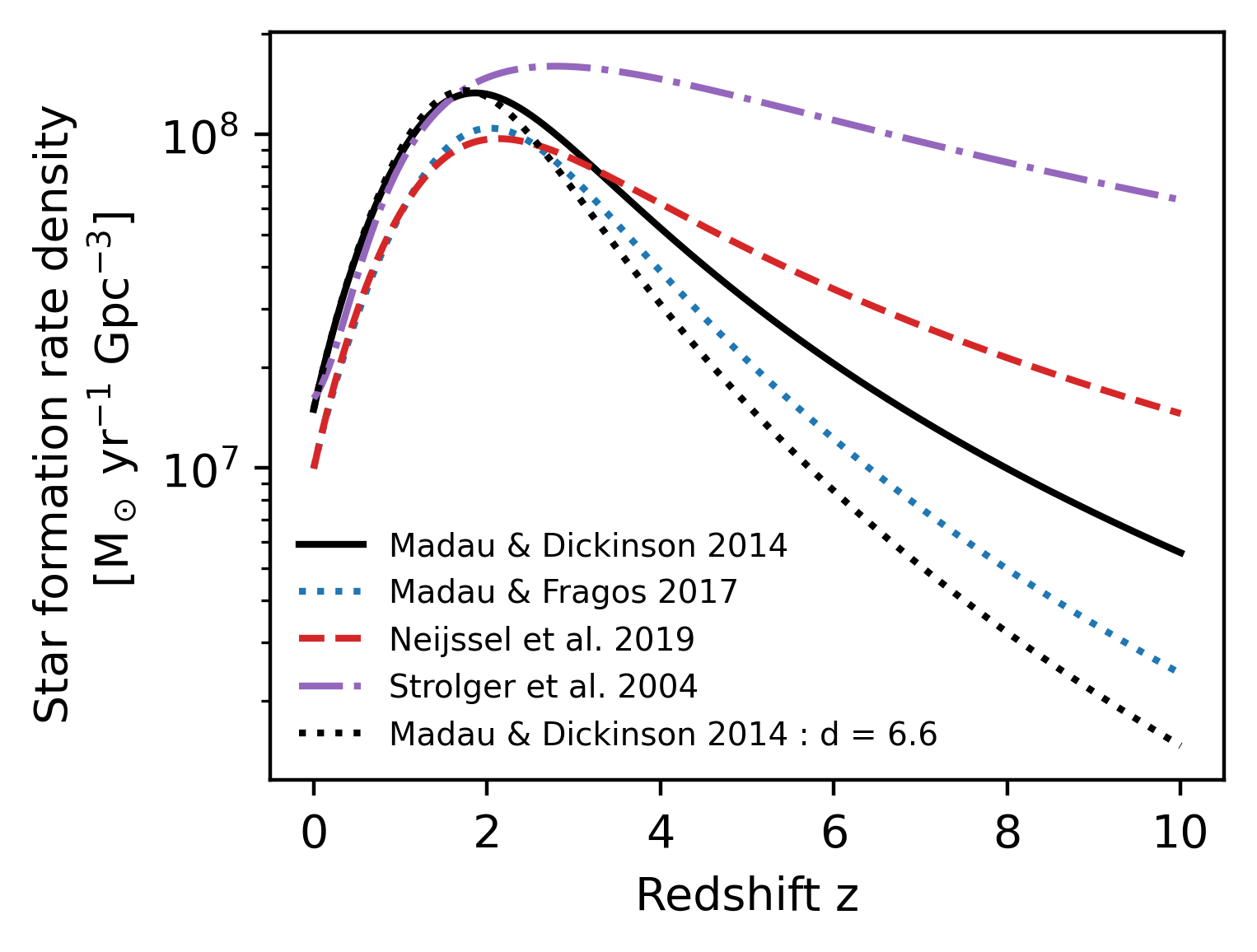}
    \caption{
    Star formation rate density as a function of redshift. 
    The solid black line shows the model from \citet{Madau:2014bja}; the dashed black lines uses their default values for $a$, $b$ and $c$ (see Equation~\ref{eq:SFR}), whilst using either $d = 3.6$ (upper curve) or $d = 6.6$ (lower curve), indicating the range of our prior on $d$ (see Table~\ref{tab:hyperparameter_ranges}). 
    We also show the fit using the parameters from \citet{Madau:2016jbv} as the dotted blue curve.
    The dot-dashed purple curve uses the star formation rate model 1 from \citet{Strolger:2004ApJ}.
    The dashed red curve shows the phenomenological model from \citet{Neijssel:2019}.
    }
    \label{fig:compare_SFR}
\end{figure}

As can be seen from Equation~\ref{eq:SFR}, when formulated this way, the SFR has 4 free parameters.
To limit the parameter space we only consider variations in $d$, which we consider to be the most uncertain parameter, as it controls the star formation rate at high redshift in this model, which is where the star formation rate is most uncertain. 
Based on the literature discussed above, we find that varying $d$ between 3.5 and 7.0 covers the range of high-redshift star formation rates discussed by the authors above.
We show a range of prescriptions for the star formations rate as a function of redshift in Figure~\ref{fig:compare_SFR}.

\subsubsection{Metallicitiy specific star formation history}
\label{subsubsec:MSSFR}

In addition to the total star formation rate at a given redshift, the distribution of star forming mass with gas phase metallicity is also a crucially important ingredient in the formation of compact object binaries (and binary black holes in particular) since both the evolution of massive stars and the formation of black holes are sensitive to the stellar metallicity (as discussed in Section~\ref{subsec:winds}).

We refer to this quantity as the metallicity-specific star formation rate, given by
\begin{equation}
    \Phi(Z,z) = \psi(z) \frac{\rm{d} N}{\rm{d} Z} ,
    \label{eq:MSSFR}
\end{equation}
where $\psi$ is the total star formation rate given by Equation~\ref{eq:SFR} and $\rm{d} N / \rm{d} Z$ is the distribution of metallicities.

In COMPAS we make use of a phenomenological gas-phase metallicity distribution, which is modelled as a log-normal distribution with a mean metallicity $\langle Z \rangle$, with a standard deviation $\sigma_Z$ \citep[][]{Neijssel:2019,COMPAS:2021methodsPaper}, independent of redshift. 
Our default model assumes that $\sigma_Z = 0.5$ \citep{Neijssel:2019}.
In our model, the mean metallicity $\langle Z \rangle$ scales with redshift $z$ as
\begin{equation}
    \langle Z \rangle = \mu_{0} \times 10^{m_\mathrm{z} z} ,
    \label{eq:mean_metallicity_redshift}
\end{equation}
or equivalently
\begin{equation}
    \log_{10} \langle Z / Z_\odot \rangle = \log_{10} \left( \frac{\mu_0}{Z_\odot} \right) + m_z z ,
    \label{eq:log_mean_metallicity_over_solar_redshift}
\end{equation}
where $\mu_0$ is the mean metallicity at redshift $z=0$, $Z_\odot$ is the solar metallicity and $m_\mathrm{z}$ describes the scaling with redshift, following \citet{Langer:2005hu}.
\citet{Langer:2005hu} assumed that $\mu_0 = Z_\odot$ and $m_z = 0.15$, based upon observationally determined metallicitities for star forming galaxies at $z < 1$ \citep{Kewley:2005}. 
A more recent study expanding the sample of galaxies (and the range of redshifts) by \citet{Yuan:2013ApJ} finds overall similar results.

Based on observations, \citet{Madau:2016jbv} assume that the mean metallicity as a function of redshift is given by $\log \langle Z / Z_\odot \rangle = 0.153 - 0.074 z^{1.34}$.
There is no set of parameters for Equation~\ref{eq:mean_metallicity_redshift} that would exactly reproduce the fit from \citet{Madau:2016jbv}.
\citet{Madau:2016jbv} find that $\log_{10} \langle Z_0 / Z_\odot \rangle = 0.153$, implying that the average star formation in the local universe is super-solar.

\begin{figure}
    \centering
    \includegraphics[width=\columnwidth]{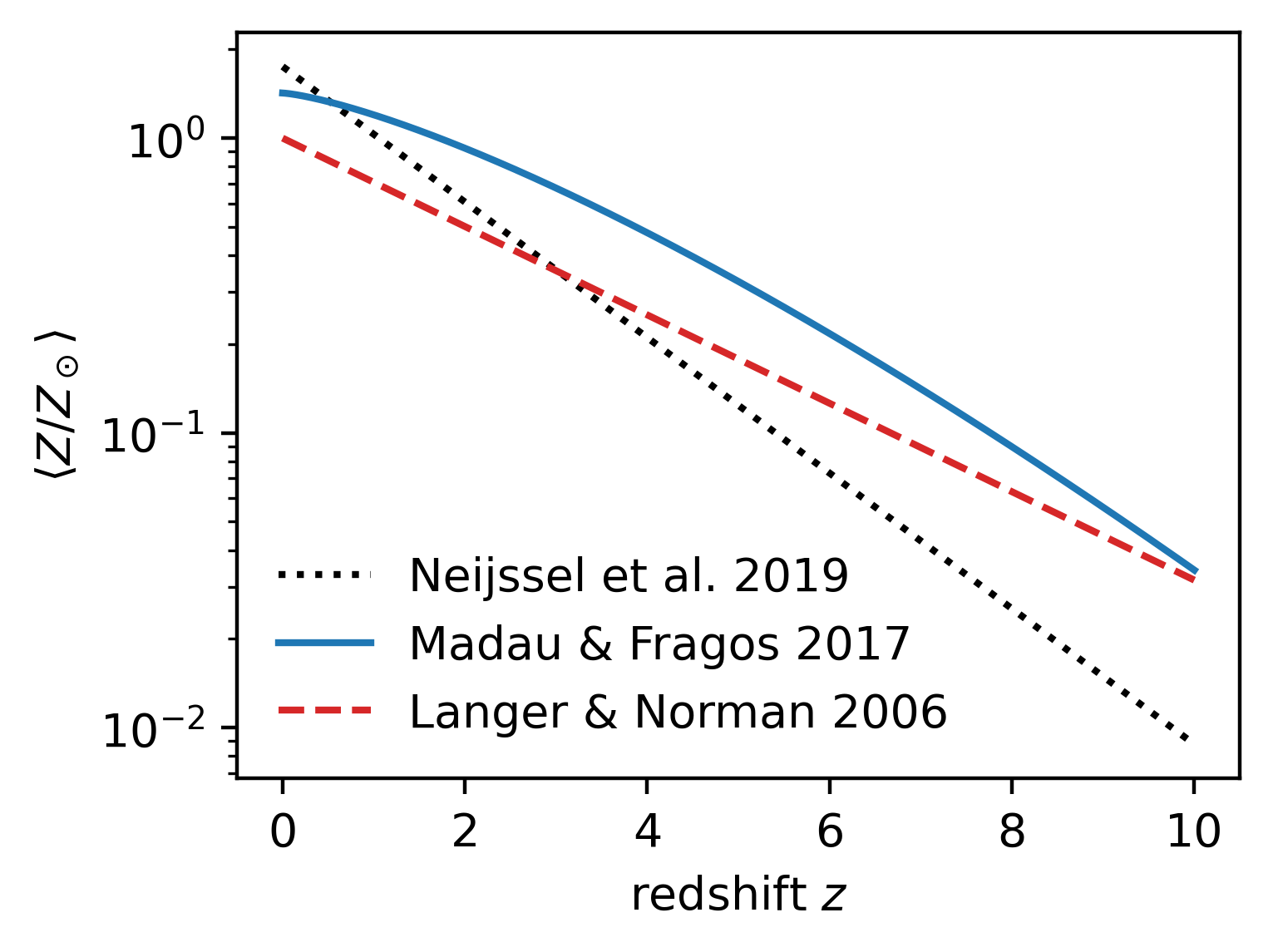} 
    \caption{Mean metallicity of gas-phase star forming material as a function of redshift as determined by the models discussed in Section~\ref{subsubsec:MSSFR}. 
    The solid blue line shows the mean metallicity as given by the fit from \citet{Madau:2016jbv}.
    The dashed red line corresponds to Equation~\ref{eq:mean_metallicity_redshift} with values taken from \citet{Langer:2005hu}, whilst the dotted black line uses values from \citet{Neijssel:2019}.}
    \label{fig:log_mean_metallicity_redshift}
\end{figure}

\citet{Neijssel:2019} found that $\mu_0 = 0.035$ and $m_\mathrm{z} = -0.23$, combined with the default binary evolution parameters in COMPAS, gave a good agreement with the observed binary black hole mass distribution and rate. 
As mentioned above, this did not include the contribution of the chemically homogeneous evolution channel and thus the best fit parameters for Equation~\ref{eq:mean_metallicity_redshift} may also change.
We show the evolution of the mean metallicity with redshift, as determined by these three prescriptions, in Figure~\ref{fig:log_mean_metallicity_redshift}. 
We use the variation between these prescriptions to motivate the range of values of $Z_0$ that we explore. 
We find a range of $Z_0$ between 0.01 and 0.05 captures the uncertainty in the average metallicity of star forming gas at redshift 0 (Table~\ref{tab:hyperparameter_ranges}).

The evolution of the metallicity specific star formation rate with redshift is highly uncertain \citep[][]{Chruslinska:2018hrb,Chruslinska:2019MNRAS}, and uncertainties in this quantity translate to large uncertainties in predictions for the rates and properties of binary black hole mergers \citep[][]{Neijssel:2019,Belczynski:2017gds,Tang:2019qhn,Boco:2020pgp,Santoliquido:2020axb,Broekgaarden:2021efa}.

In addition to the uncertainties related to the average star formation rate at redshift 0, there are also uncertainties related to the evolution of the average metallicity with redshift (as characterised by $m_Z$) and the distribution of metallicities at a given redshift, as characterised by $\sigma_\mathrm{Z}$, which may or may not be redshift-dependent. 
We leave a more thorough exploration of this problem for future work.

\subsection{Merger rates}
\label{subsec:merger_rates}

One of the key observables predicted by our binary population synthesis models is the rate of binary black hole mergers.
In this section we define several different rates that we quote in this work.

We calculate the differential merger rate of binary black holes at redshift $z$ per unit volume per unit time as a function of chirp mass $\mathcal{M}$ according to
\begin{equation}
     \frac{\rm{d} N}{\rm{d} V_\mathrm{c} \rm{d} t_\mathrm{s} \rm{d} \mathcal{M} } (z) =  
     \int \rm{d} Z \int \rm{d} \tau_\mathrm{delay}
     \mathcal{R}_\mathrm{form}
     \Phi(Z, t_\mathrm{form}) ,
    \label{eq:local_rate_func_redshift}
\end{equation}
where $t_\mathrm{s}$ is the time measured in the source frame of the merging binary black hole, and $V_\mathrm{c}$ is the comoving volume. The term $\Phi(Z, t_\mathrm{form}$) in Equation~\ref{eq:local_rate_func_redshift} is the metallicity-specific star formation rate, given by Equation~\ref{eq:MSSFR}.

The binary chirp mass $\mathcal{M}$ is a combination of the component masses $m_{1}$ and $m_{2}$ given by 
\begin{equation}
    \mathcal{M} = \frac{(m_1 m_2)^{3/5}}{(m_1 + m_2)^{1/5}} .
    \label{eq:chirp_mass}
\end{equation}

The term $\mathcal{R}_\mathrm{form}$ in Equation~\ref{eq:local_rate_func_redshift} describes the formation rate of binary black holes estimated by COMPAS, and is given by
\begin{equation}
    \mathcal{R}_\mathrm{form} = 
    \frac{\rm{d N}}{\rm{d} M_\mathrm{form} \rm{d} \tau_\mathrm{delay} \rm{d} \mathcal{M} } ,
    \label{eq:formation_efficiency}
\end{equation}
where $M_\mathrm{form}$ is the amount of star forming mass and $\tau_\mathrm{delay}$ is the delay time between star formation and the compact object merger \citep{Peters:1964}. The time at which the progenitor stars form is related to the time the binary merges $t_\mathrm{merge}$ by $t_\mathrm{form} = t_\mathrm{merge}(z) - \tau_\mathrm{delay}$ \citep[e.g.,][]{Barrett:2017fcw}.

Where necessary, we adopt the cosmological parameters for a flat $\Lambda$-CDM universe from \citet{Hinshaw:2013ApJS} as implemented in \textsc{astropy} \citep{astropy:2013,astropy:2018,astropy:2022}. 

The merger rate at redshift $z$ is then given by integrating Equation~\ref{eq:local_rate_func_redshift} over all chirp masses
\begin{equation}
    \mathcal{R} (z) = 
    \frac{\rm{d} N}{\rm{d} V_\mathrm{c} \rm{d} t_\mathrm{s}}
    = \int \frac{\rm{d} N}{\rm{d} V_\mathrm{c} \rm{d} t_\mathrm{s} \rm{d} \mathcal{M} } (z) \rm{d} \mathcal{M} .
    \label{eq:merger_rate_at_redshift}
\end{equation}
We define the local merger rate as the value given by Equation~\ref{eq:merger_rate_at_redshift} evaluated at redshift $z = 0$.

The predicted observed chirp mass distribution differs from that given by Equation~\ref{eq:local_rate_func_redshift} and is given by
\begin{equation}
    \frac{\rm{d} N_\mathrm{det}}{\rm{d} t_\mathrm{obs} \rm{d} \mathcal{M}} = 
    \int \mathrm{d} z 
    \frac{\rm{d} N}{\rm{d} t_{s} \rm{d} V_{c} \rm{d} \mathcal{M}}
    \frac{\rm{d} V_{c}}{\mathrm{d} z}
    \frac{\rm{d} t_{s}}{\rm{d} t_\mathrm{obs}}
    P_\mathrm{det} ,
    \label{eq:predicted_observed_chirp_mass_dist}
\end{equation}
where the first term inside the integral is given by Equation~\ref{eq:local_rate_func_redshift}, $\rm{d} t_\mathrm{obs} = (1 + z) \rm{d} t_\mathrm{s}$ is the time measured in the frame of the observer, $\rm{d} V_{c}/\rm{d} z$ is the volume element\footnote{We calculate the cosmological volume element $\rm{d} V_{c}/\rm{d} z$ using \textsc{Astropy} \citep{astropy:2013,astropy:2018,astropy:2022}} and $P_\mathrm{det}$ is the detection probability for a given binary \citep{Barrett:2017fcw}.

The detection probability $P_\mathrm{det}$ for a given binary depends on the parameters of that binary, such as its masses, spins and distance, and therefore describes the selection effects imposed by gravitational-wave observatories.
We determine $P_\mathrm{det}$ using the method described by \citet{Barrett:2017fcw}. 
In summary, we use a phenomenological model \citep[\textsc{IMRPhenomPv2};][]{Khan:2016Phenom} that includes the inspiral, merger and ringdown phases of the gravitational waveform. 
When calculating sensitivities, we neglect black hole spins which are expected to play a subdominant role in the detectability of binary black holes \citep{Ng:2018neg}.
Regardless, the majority of black hole spins inferred from gravitational-wave observations appear to be small \citep{Farr:2017uvj,LIGOScientific:2020kqk,Galaudage:2021rkt}.
Using this model, we estimate the signal-to-noise ratio a given binary would produce in a single gravitational-wave observatory operating with a sensitivity comparable to that of the LIGO inteferometers during their third observing run \citep[O3;][]{LIGOScientific:2021djp,LVKObservingScenarios2020}.
Following convention, we assume that binaries that produce a signal-to-noise ratio greater than 8 are detected.

The predicted detection rate is given by integrating Equation~\ref{eq:predicted_observed_chirp_mass_dist} over all chirp masses
\begin{equation}
    \mathcal{R}_\mathrm{det} = \frac{\mathrm{d} N_\mathrm{det}}{\rm{d} t_\mathrm{obs}} = 
    \int \mathrm{d} \mathcal{M}
    \frac{\rm{d} N}{\rm{d} t_\mathrm{obs} \rm{d} \mathcal{M}} ,
    \label{eq:detection_rate}
\end{equation}
and the expected number of detections in an observing period of duration $T_\mathrm{obs}$ is given by 
\begin{equation}
    N_\mathrm{det}  = \mathcal{R}_\mathrm{det} T_\mathrm{obs} .
    \label{eq:number_of_detections}
\end{equation}

\subsection{Model exploration}
\label{subsec:model_exploration}

\begin{table*}
    \centering
    \begin{tabular}{l|c|c|c}
         Parameter & Fiducial value & Minimum & Maximum \\
         \hline \\
         Common envelope efficiency ($\alpha_\mathrm{CE}$) & 1 & 0.1 & 10 \\
         Wolf--Rayet mass-loss rate ($f_\mathrm{WR}$) & 1 & 0.1 & 10 \\
         High-redshift star-formation rate ($d$) & 4.7 & 3.6 & 6.6 \\
         Average star-formation metallicity ($Z_0$) & 0.035 & 0.01 & 0.05 \\
    \end{tabular}
    \caption{Hyperparameters explored in this paper. 
    The fiducial value refers to our default assumption for each parameter, taken from \citet{COMPAS:2021methodsPaper}. 
    We also list the minimum and maximum value that each parameter is varied between (as justified in Section~\ref{sec:methods}).}
    \label{tab:hyperparameter_ranges}
\end{table*}

As discussed in the introduction, there are many uncertainties in massive binary evolution, and ideally, population synthesis analyses would explore the full range of possibilities for each uncertainty.
However, if one wishes to explore $N_\mathrm{hyper}$ hyperparameters with $N_\mathrm{explore}$ models per hyperparameter, then the total number of models required will be roughly $N_\mathrm{explore}^{N_\mathrm{hyper}}$. This number can very quickly become large for even moderate values of  $N_\mathrm{explore}$ and $N_\mathrm{hyper}$.
This is the well known \emph{curse of dimensionality}.

In order to avoid the \emph{curse of dimensionality} we have chosen to limit the parameters we explore to the 4 parameters introduced earlier in Section~\ref{sec:methods}, namely the efficiency of common envelope evolution ($\alpha_\mathrm{CE}$), the mass-loss rates of Wolf--Rayet stars ($f_\mathrm{WR}$), the cosmic star formation rate at high redshift ($d$) and the average metallicity of star formation at redshift 0 ($Z_0$), which we term \emph{population hyperparameters} \citep[][]{Barrett:2017fcw}.
We have chosen these parameters as previous work has shown that they have the largest impact on predictions for merging binary black holes \citep[e.g.,][]{Riley:2020btf,Broekgaarden:2021efa}

We construct our grid of models using Latin hypercube sampling \citep{McKay:LHS} as implemented in the Python package \textsc{PyDOE}. 
We use an algorithm that maximises the minimum distance between points \citep[][]{Morris:1995}.
This distribution has the property that the marginalised one-dimensional distributions are uniform in the sampled parameter.
We draw 50 samples using the Latin hypercube sampling, and additionally include the edges of the parameter space. 
In total we sample $N = 54$ models with different combinations of the binary evolution parameters $\alpha_\mathrm{CE}$ and $f_\mathrm{WR}$.

\begin{figure}
    \centering
    \includegraphics[width=\columnwidth]{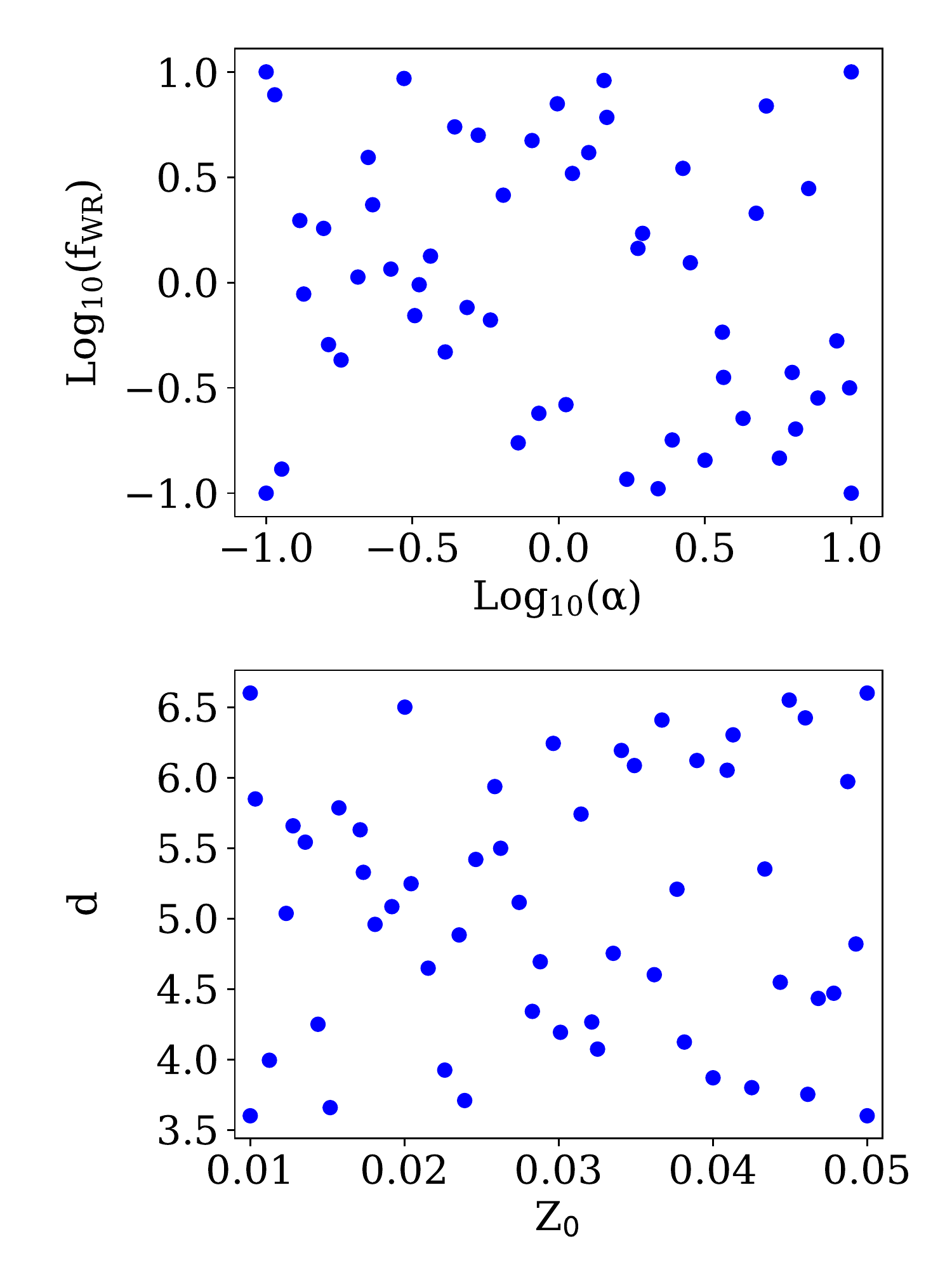}
    \caption{Hyperparameters at which we calculate our population synthesis models, drawn using Latin hypercube sampling. The top panel shows the binary evolution parameters $\alpha$ and $f_\mathrm{WR}$, whilst the bottom panel shows the parameters governing the cosmic star formation history, $d$ and $Z_{0}$.}
    \label{fig:hyperparameters}
\end{figure}

Since they are decoupled from the underlying binary evolution, we explore the impact of uncertainties in the cosmic star formation history in post-processing. 
We use $N = 54$ combinations of SFR parameters for each combination of binary evolution parameters.
Thus in total we explore $54 \times 54 = 2,916$ different models.
We list the ranges of parameters we explore in Table~\ref{tab:hyperparameter_ranges}.
We draw $\alpha_\mathrm{CE}$ by sampling $\log_{10} (\alpha_\mathrm{CE})$ between $\log_{10} (\alpha_\mathrm{CE}^\mathrm{min})$ and $\log_{10} (\alpha_\mathrm{CE}^\mathrm{max})$, as specified in Table~\ref{tab:hyperparameter_ranges}.
Similarly, for the mass-loss rates of Wolf--Rayet stars, we determine the multiplier $f_\mathrm{WR}$ by drawing $\log_{10} (f_\mathrm{WR})$ between $\log_{10} (f_\mathrm{WR}^\mathrm{min})$ and $\log_{10} (f_\mathrm{WR}^\mathrm{max})$.
We draw the values of $d$ and $Z_0$ between $d^\mathrm{min}$ and $d^\mathrm{max}$, and $Z_0^\mathrm{min}$ and $Z_0^\mathrm{max}$ respectively.
We show these samples in Figure~\ref{fig:hyperparameters}.

For each model, we sample the metallicities of the binaries uniformly in the log between $Z_\mathrm{min} = 10^{-4}$ and $Z_\mathrm{max} = 0.03$ \citep{COMPAS:2021methodsPaper}. 
This helps to mitigate issues associated with using grids of discrete metallicities, such as sharp features (`spikes') in the mass distribution of binary black holes \citep[][]{Dominik:2014yma}.
We draw $N = 10^{6}$ binaries from each model.

Our models typically take $\sim 12$\,hrs to compute on a single node on the OzSTAR supercomputer at Swinburne University of Technology.
We make use of the trivial parallelisability of COMPAS in order to run multiple models in parallel, utilising up to 54 cores.
We find that post-processing contributes only a negligible additional computational cost.
More sophisticated sampling techniques could reduce the computational cost of exploring the parameter space of binary evolution, or equivalently, either allow one to broaden the parameter space explored, or decrease statistical uncertainties at a fixed computational cost \citep{Broekgaarden:2019qnw}.

\section{Observational sample}
\label{sec:observational_sample}

\begin{figure}
    \centering
    \includegraphics[width=0.9\columnwidth]{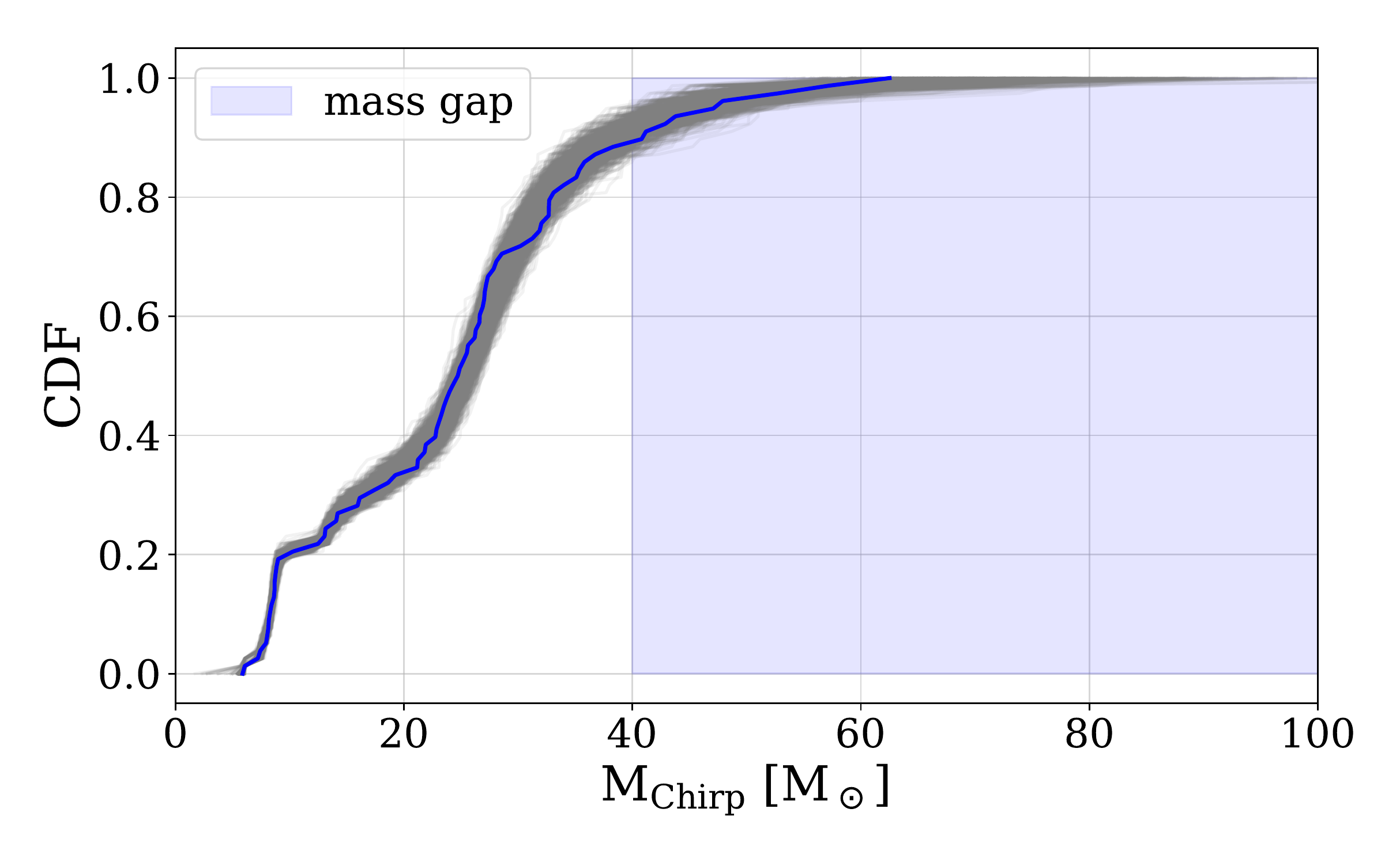}
    \includegraphics[width=0.9\columnwidth]{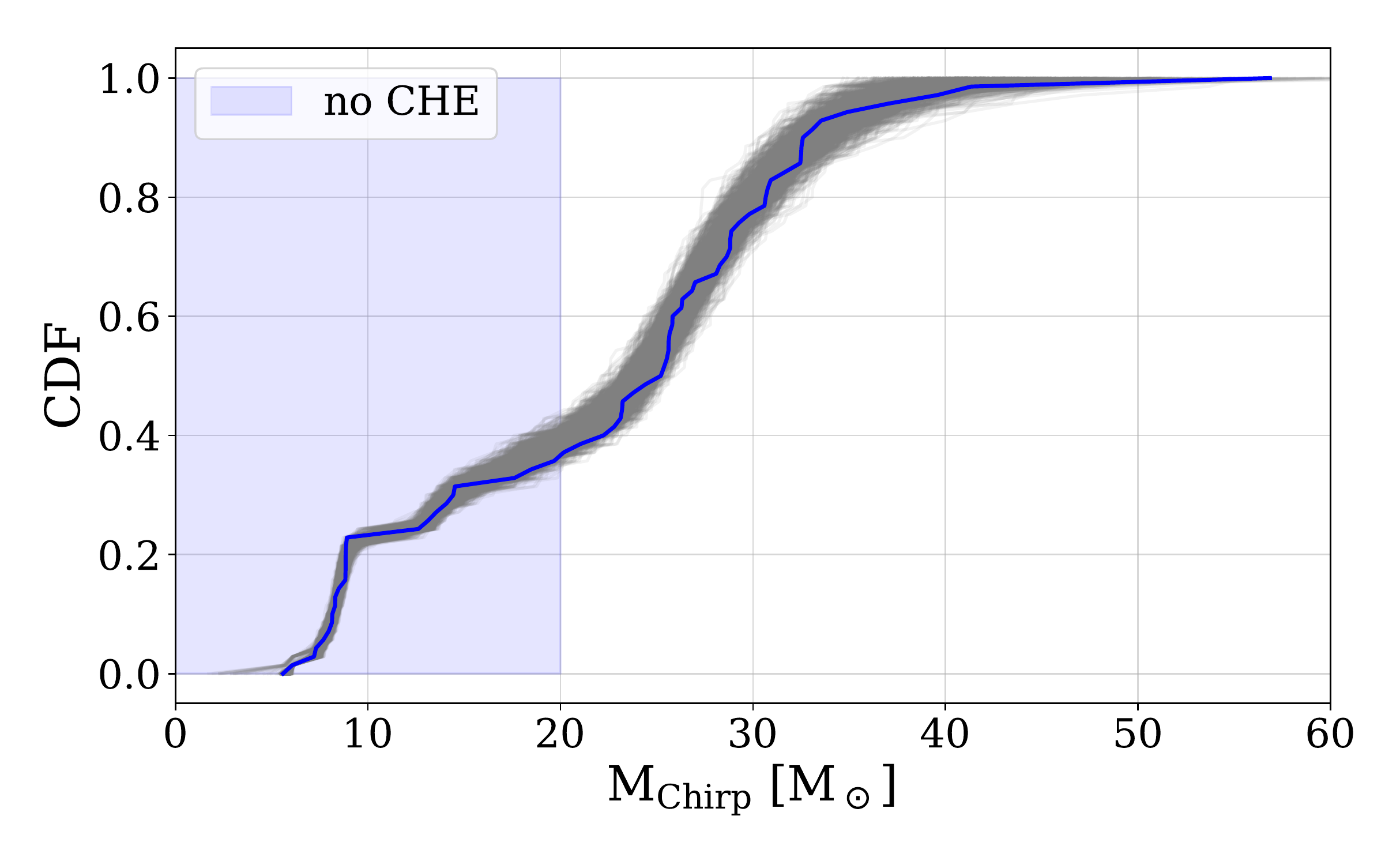}
    \includegraphics[width=0.9\columnwidth]{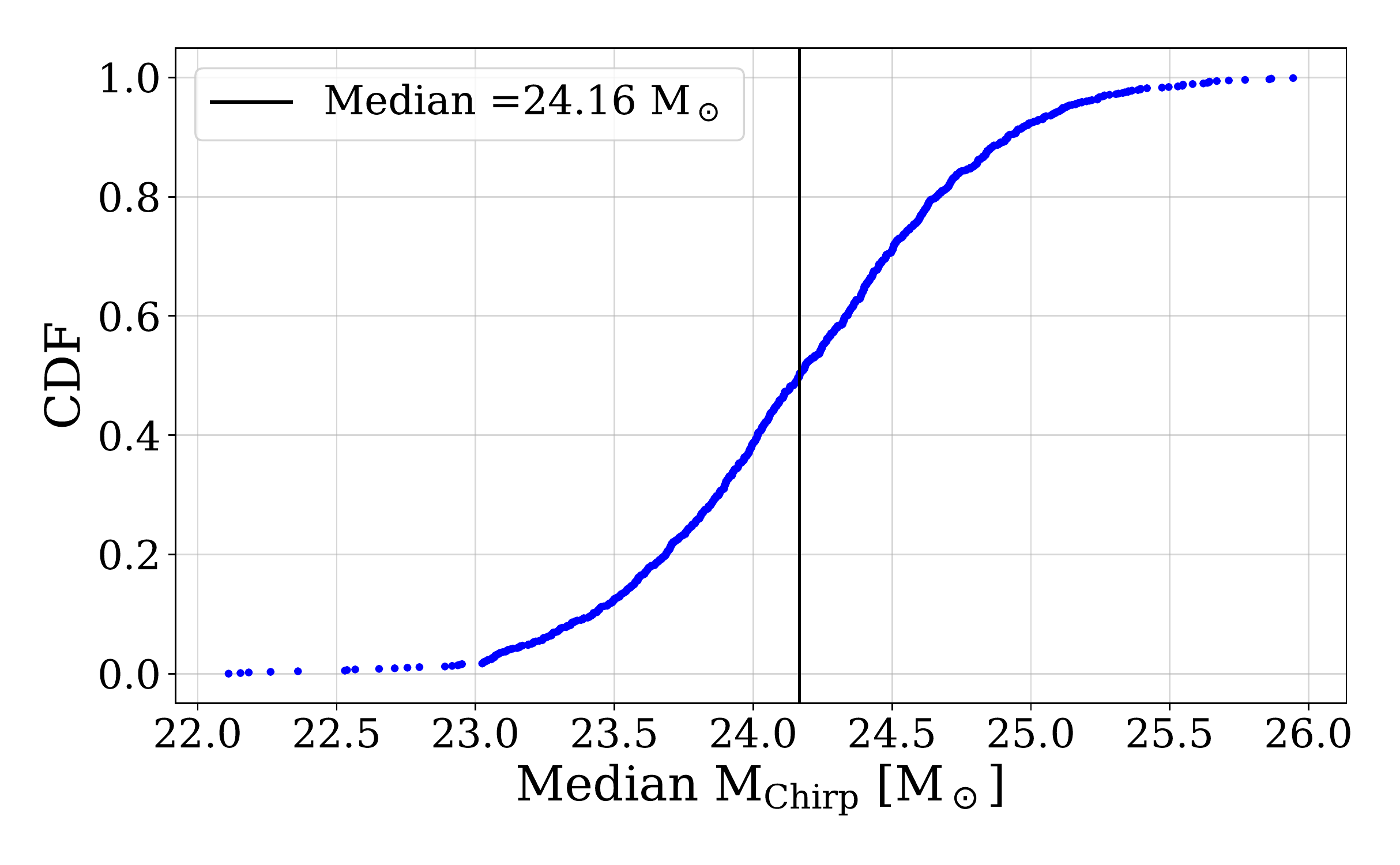}
    \caption{Empirical cumulative chirp mass distribution of binary black holes detected during the first three observing runs of Advanced LIGO and Virgo \citep{LIGOScientific:2018mvr,Abbott:2021PhRvXGWTC-2,LIGOScientific:2021djp}. 
    In the top panel, each of the 1000 lines is constructed by randomly drawing one sample from the posterior distribution for the chirp mass for each of the 79 binary black hole events observed in gravitational waves. 
    The 10\% of binary black holes with chirp masses greater than 40\,M$_\odot$ (shaded blue region) are unlikely to have formed through isolated binary evolution (see Section~\ref{sec:observational_sample} for more discussion).
    In the middle panel we exclude all binary black holes with a median chirp mass greater than 40\,M$_\odot$. 
    The solid blue lines in the top two panels highlight a single random cumulative distribution to guide the eye.
    The shaded blue region shows that of the binary black holes that can be formed through binary evolution, $\sim 40\%$ have masses less than the lowest mass produced through chemically homogeneous evolution. 
    The bottom panel shows the cumulative distribution for the median observed chirp mass, which is around $25$\,M$_\odot$.
}
    \label{fig:observed_chirp_mass_dist}
\end{figure}

\begin{figure}
	\includegraphics[width=0.9\columnwidth]{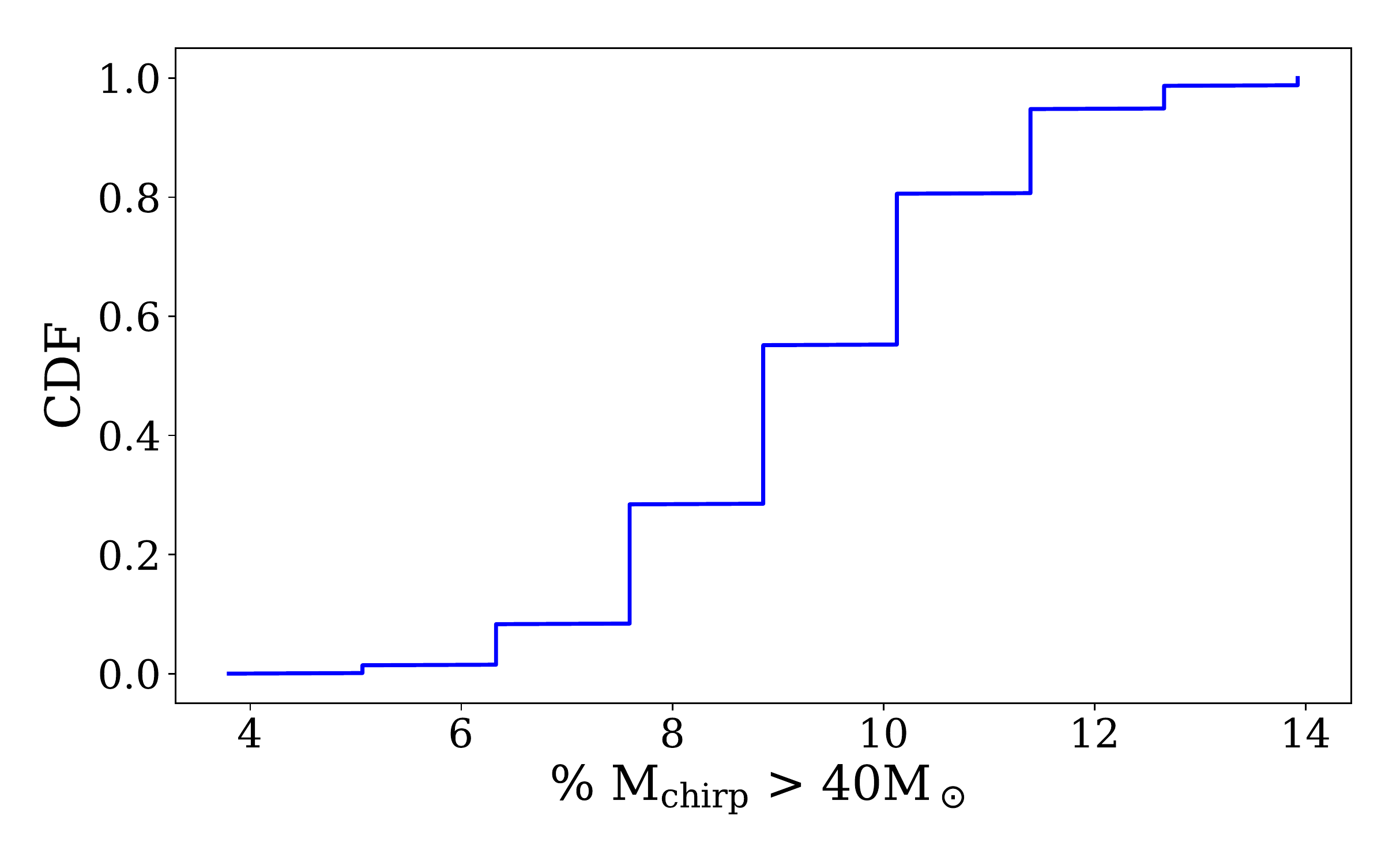}
    \caption{Cumulative distribution of the percentage of observed binary black holes with chirp masses greater than $40$\,M$_\odot$, the maximum obtained in COMPAS \citep[][]{Stevenson:2019rcw}.
    The histogram is generated from 1000 posterior samples for each of the 79 BBH events in GWTC-3 \citep[][]{LIGOScientific:2021djp}. 
    Approximately nine percent of the observed BBH population have a chirp mass that cannot be reproduced in COMPAS.}
    \label{fig:frac_too_massive}
\end{figure}

Before comparing our models to the gravitational-wave observations, we first discuss the details of our sample selection.
We start from all gravitational-wave events included in the catalogue of gravitational-wave transients published following LIGO and Virgo's third observing run \citep[GWTC-3;][]{LIGOScientific:2021djp}.
This includes all events that have an estimated probability of astrophysical origin, $P_\mathrm{astro}$, greater than $0.5$ \citep{LIGOScientific:2021djp}.
We begin by excluding all events in which the secondary has a mass $m_2$ consistent with a neutron star, since these could either be binary neutron star (such as GW170817 and GW190425; \citealp{LIGOScientific:2017vwq,LIGOScientific:2020aai}) or neutron star-black hole (such as GW200105; \citealp{LIGOScientific:2021qlt}) binaries.
We also choose to exclude the highly asymmetric binary GW190814 \citep{LIGOScientific:2020zkf}, partly due to the uncertain classification of the secondary (which may either be the most massive neutron star or lightest black hole observed), and partly due to the mass ratio, which is difficult to produce though isolated binary evolution \citep[though see][for possible explanations in this context]{Zevin:2020gma,Mandel:2020cig,Antoniadis:2021dhe}.
We assume that the remainder of the population ($79$ events) are binary black holes.

As discussed in Section~\ref{sec:methods}, binary evolution models (including COMPAS) predict the existence of a gap in the mass spectrum of black holes, beginning above $\sim 45$\,M$_\odot$ \citep[e.g.,][]{Stevenson:2019rcw}, although uncertainties (mainly in nuclear reaction rates) may allow this value to be higher \citep[][]{Farmer:2019ApJ}.
We find that the maximum binary black hole chirp mass does not vary significantly between our models.
We find a maximum chirp mass of around $40$\,M$_\odot$ in our models (see Section~\ref{subsec:BBH_mass_dist}).
We therefore choose to exclude all observed binary black holes with chirp masses greater than this value, as in our analysis, these events must be produced via other formation pathways such as hierarchical mergers of lower mass black holes through stellar dynamics \citep[e.g.,][]{Rodriguez:2017pec,Yang:2019cbr,DiCarlo:2019fcq,Mapelli:2020xeq,Mapelli:2021syv}.
This excludes events such as GW190521 \citep{Abbott:2020tfl,Abbott:2020mjq} and GW200220\_061928 \citep[][]{LIGOScientific:2021djp}.
Applying these selection criteria leaves us with a sample of $71$ binary black hole observations.
We plot the empirical cumulative distribution for the chirp masses of all observed binary black holes in the top panel of Figure~\ref{fig:observed_chirp_mass_dist}, whilst we show only those events with a median chirp mass of less than $40$\,M$_\odot$ in the middle panel of Figure~\ref{fig:observed_chirp_mass_dist}.
We show the fraction of binary black holes with a chirp mass greater than $40$\,M$_\odot$ in Figure~\ref{fig:frac_too_massive}.

Intriguingly, we find that around $10\%$ of binary black holes are excluded from our comparison as they have masses greater than that predicted in our models (Figure~\ref{fig:frac_too_massive}). 
This immediately implies that at least $10\%$ of the observed binary black hole population is not formed through isolated binary evolution; we treat this fraction as a lower limit, since if there are contributions from other channels above a chirp mass of 40\,M$_\odot$, it is likely that there are contributions/events below this chirp mass limit too.

Some authors have come to similar conclusions based on alternate methods and lines of reasoning.
By searching for signatures of eccentricity in the observed gravitational waveforms, \citet{Romero-Shaw:2021ual,Romero-Shaw:2022xko} argue that up to $100\%$ of binary black holes could be dynamically formed based on the observation of significant eccentricity in 4 binaries at the time of merger (which is not expected through isolated binary evolution), including the massive binary black hole merger GW190521 \citep[][]{Romero-Shaw:2020thy,Bustillo:2020ukp,Gayathri:2022NatureAstronomy}.
Using a combination of population synthesis models, \citet{Zevin:2020gbd} find that field channels and dynamical channels may contribute similar fractions of the observed binary black hole population, whilst \citet{Wong:2020ise} use a mixture of predictions from isolated binary evolution and dynamical formation in globular clusters and find that less than half of the observed binary black holes formed through isolated binary evolution in their models. 
\citet{Bouffanais:2019ApJ} and \citet{Bouffanais:2021wcr} find support for contributions from both isolated binary evolution and dynamical formation (in their case, in young star clusters).
\citet{Safarzadeh:2020jsc}, modelling the spin distribution of binary black holes, argue that more than half of the population should arise from dynamical encounters, whereas \citet{Tauris:2022arXiv} claims that the spin distribution of binary black holes can be explained due to the reorientation of black hole spins at formation.
It seems that, despite using a range of different methods and models, the current consensus in the literature is that there are likely contributions from multiple formation channels to the observed binary black hole population.

Another interesting constraint can be obtained directly from Figure~\ref{fig:observed_chirp_mass_dist}.
We see that approximately $40$\% of observed binary black holes that can have formed through isolated binary evolution (as discussed above) have chirp masses less than $20$\,M$_\odot$.
Binary black holes formed through chemically homogeneous evolution are predominantly expected to be formed at low metallicity and have chirp masses greater than $20$\,M$_\odot$ \citep[][]{Mandel:2015qlu,deMink:2016vkw}.
Therefore, around $40$\% of binary black holes have chirp masses which are too low to have formed through the chemically homogeneous evolution channel (see Section~\ref{subsec:BBH_mass_dist}). 
This therefore places a lower limit on the fraction of observed binary black holes formed through other channels.
Or alternatively, one can view this as an upper limit of around $60$\% on the fraction of all detected binary black holes that may have formed through chemically homogeneous evolution.
We discuss the uncertainties in the chemically homogeneous evolution channel further in Section~\ref{sec:conclusions}.

\section{Model predictions}
\label{sec:results}

\subsection{Binary black hole merger rates}
\label{subsec:BBH_rates}

\begin{figure}
    \centering
    \includegraphics[width=\columnwidth]{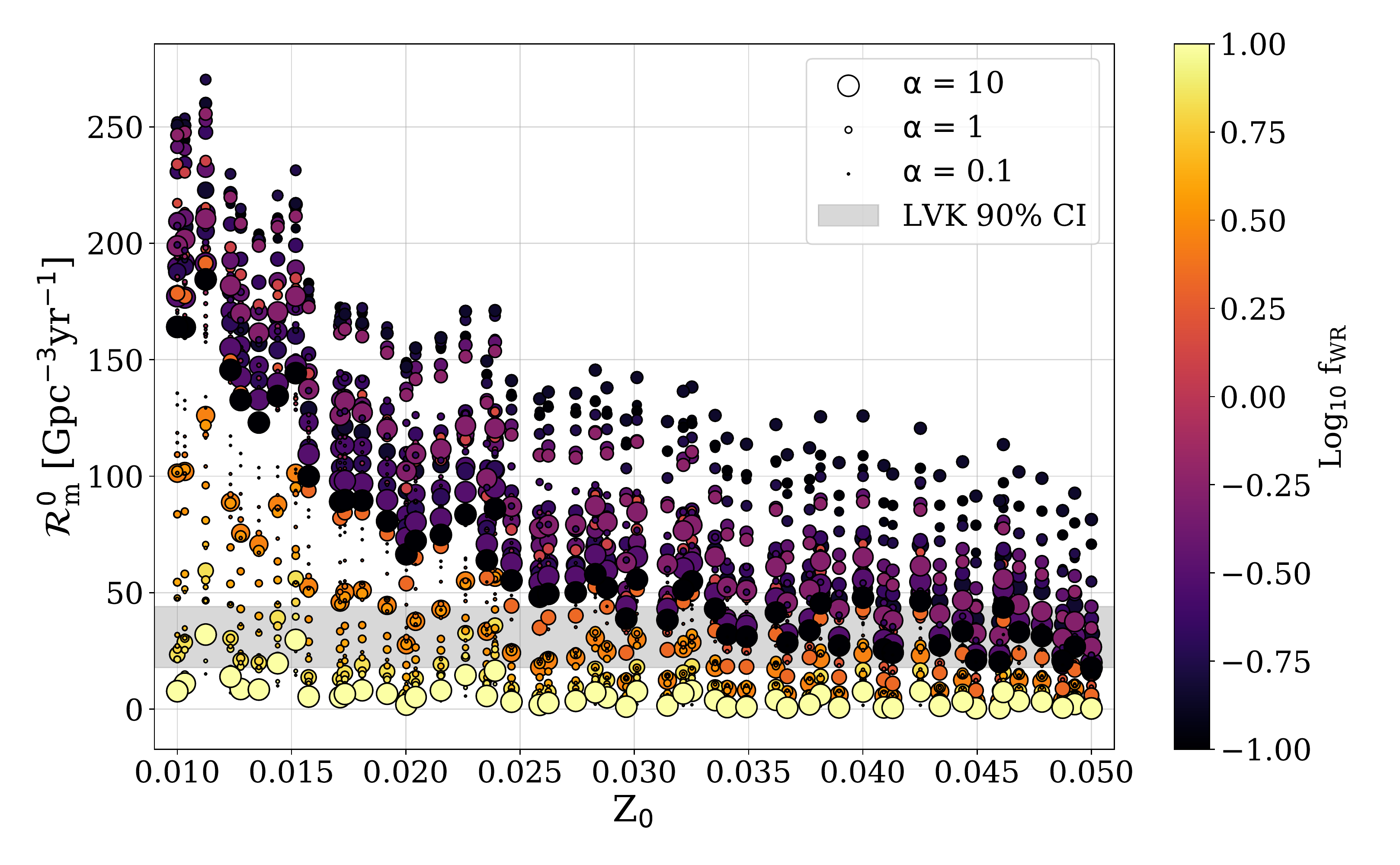}
    \includegraphics[width=\columnwidth]{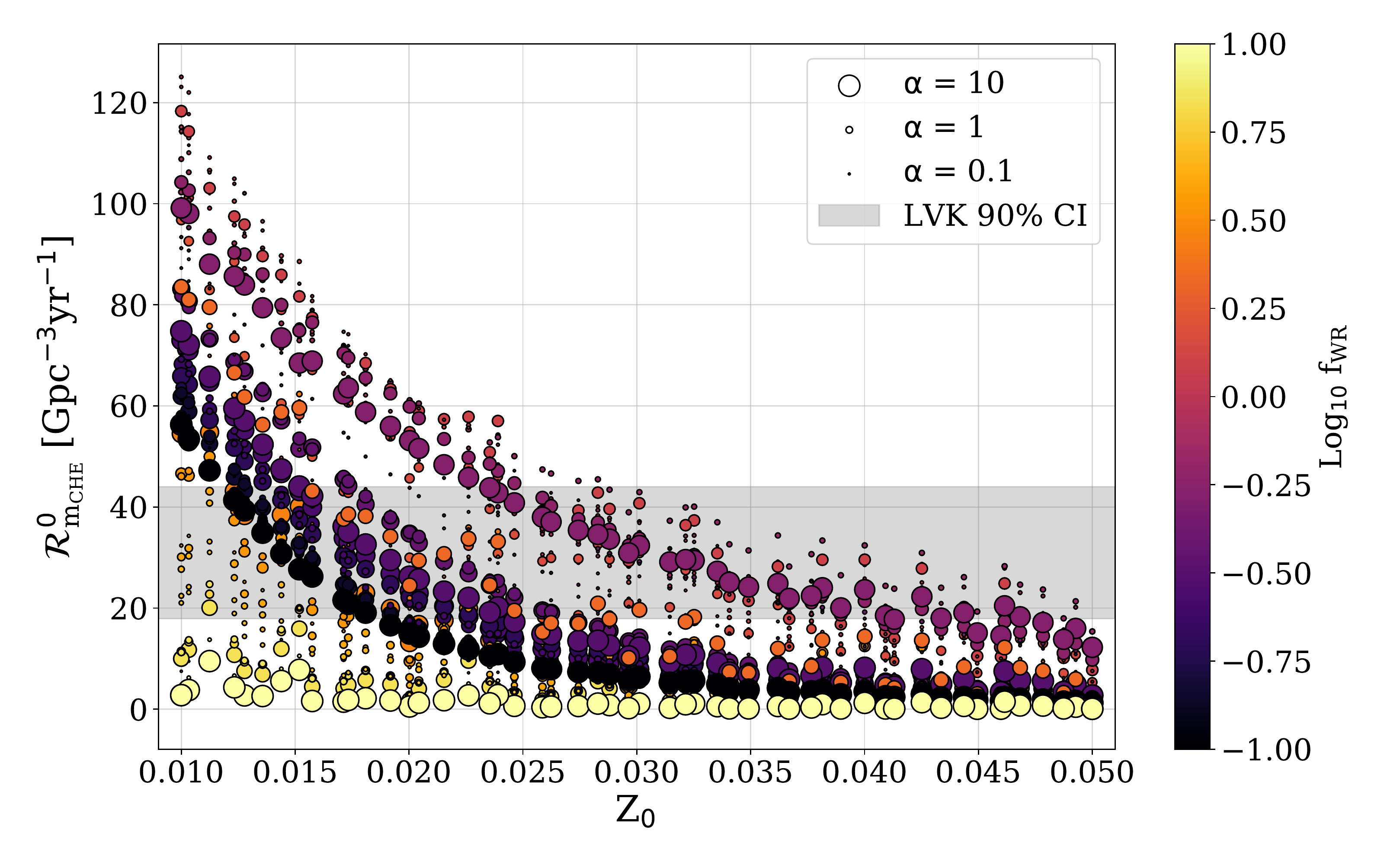}
    \caption{Predicted binary black hole merger rate at redshift $z = 0$ from our models as a function of model hyperparameters. The top panel shows the total predicted binary black hole merger rate from all formation channels, whilst the bottom panel shows only the contribution from the chemically homogeneous evolution channel. The shaded grey region in both panels shows the range of intrinsic binary black hole merger rates inferred by \citet{LIGOScientific:2021psn}.
    }
    \label{fig:BBH_rate}
\end{figure}

We begin by showing the intrinsic binary black hole merger rate at redshift $z = 0$, as determined from Equation~\ref{eq:merger_rate_at_redshift}, in Figure~\ref{fig:BBH_rate} as a function of our model hyperparameters.
We find a large range of merger rates, ranging from 10--400\,Gpc$^{-3}$\,yr$^{-1}$.
This emphasises just how much variation is possible within these models, given the current level of uncertainties (see also \citealt{Broekgaarden:2021efa} and \citealt{Mandel:2022LRR}).

\begin{figure}
    \centering
    \includegraphics[width=\columnwidth]{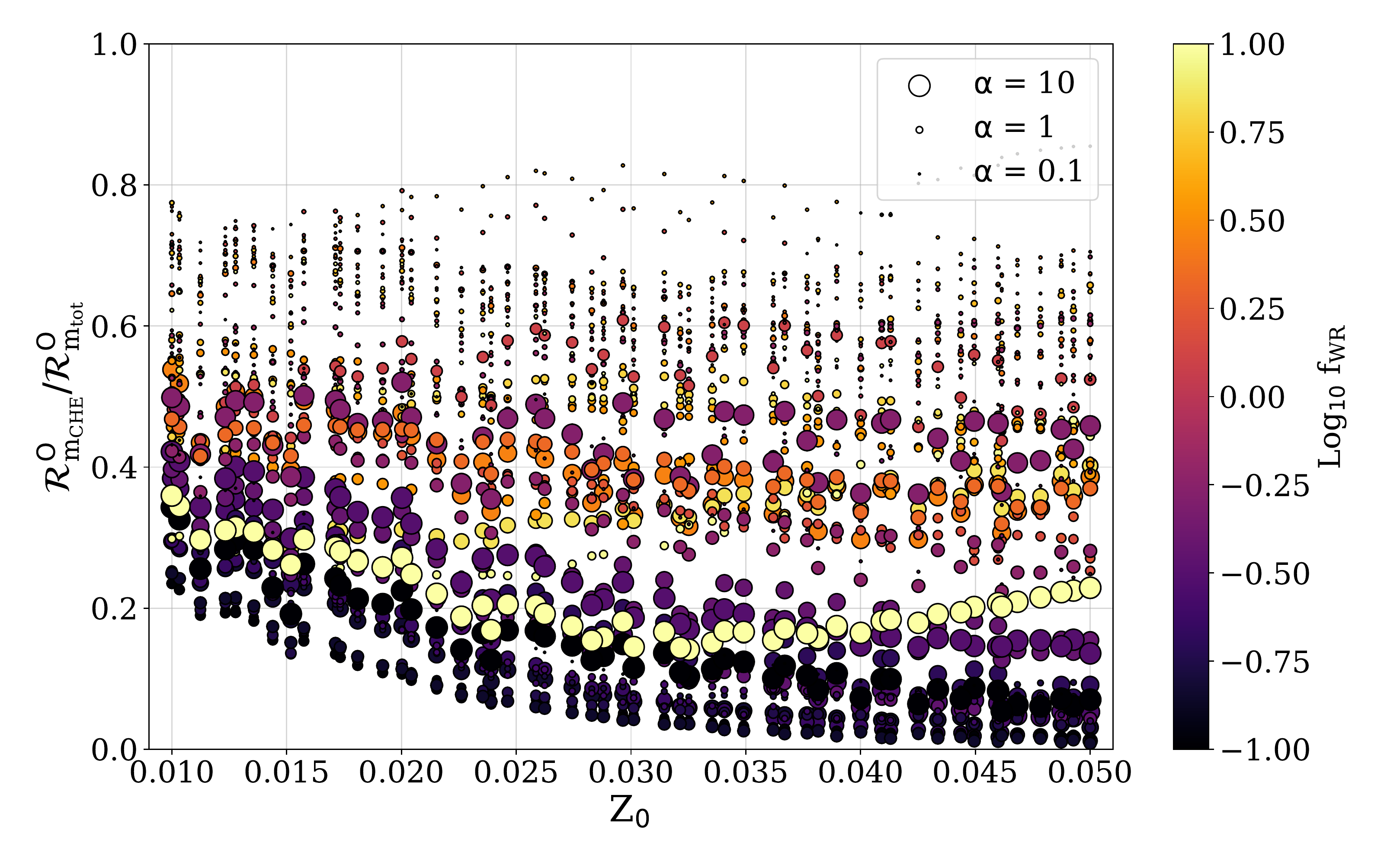}
    \includegraphics[width=\columnwidth]{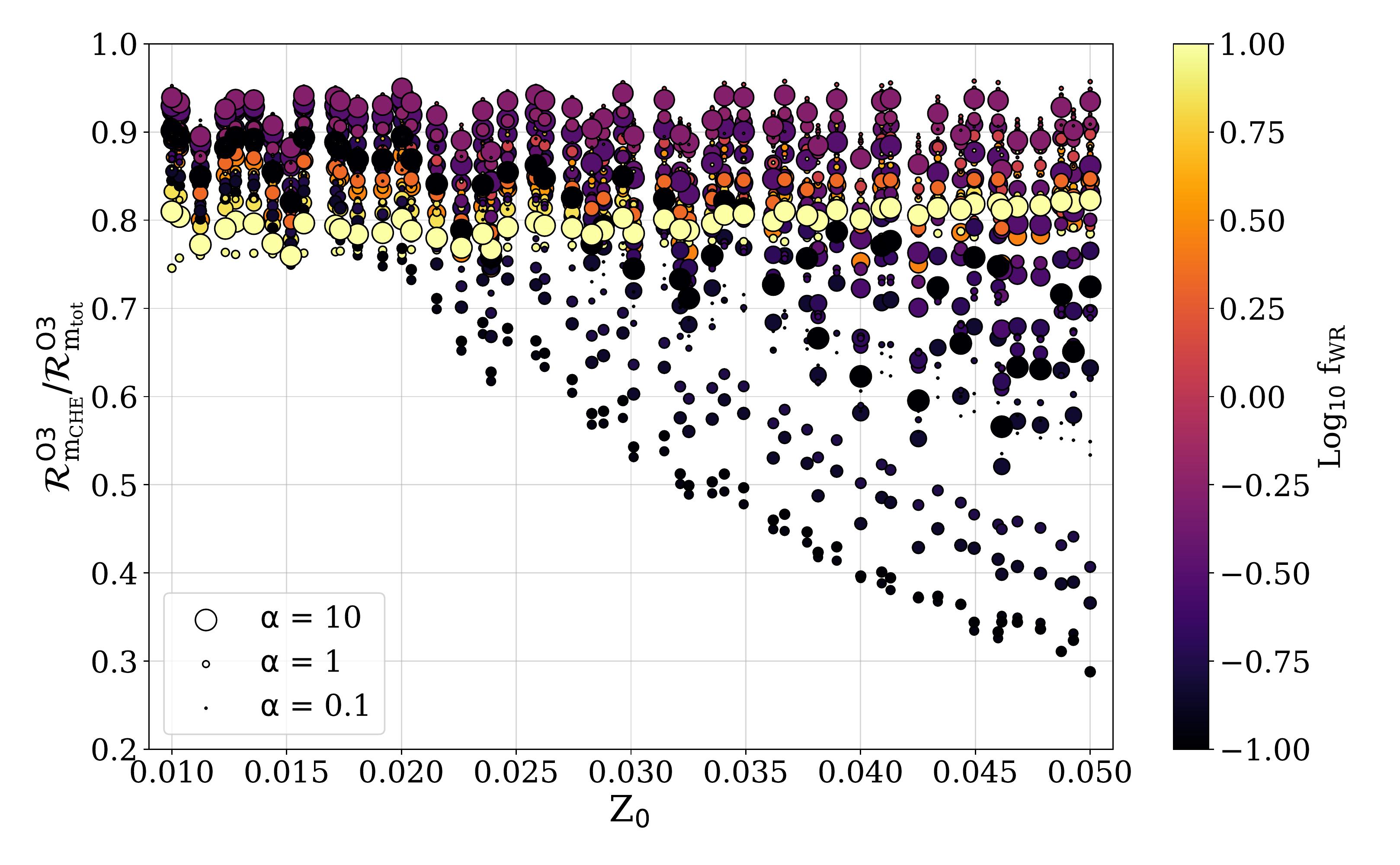}
    \caption{Fraction of merging binary black holes formed through the chemically homogeneous evolution channel, shown as a function of model hyperparameters.
    The top panel shows the intrinsic fraction, whilst the bottom panel shows the observed fraction accounting for selection effects.}
    \label{fig:fraction_CHE}
\end{figure}

The top panel of Figure~\ref{fig:BBH_rate} shows the total intrinsic binary black hole merger rate at redshift $z = 0$ from all isolated binary evolution pathways modelled within COMPAS.
We can identify several trends.
We find that models with reduced Wolf--Rayet mass-loss rates tend to predict larger rates than models with higher mass-loss rates.
Increased mass-loss during the Wolf--Rayet stage impacts the binary black hole merger rate in a couple of ways. 
Increased mass-loss trivially leads to collapsing stars being less massive.
This can lead to a larger fraction of binaries being disrupted by the natal kick given to the black holes at birth (particularly for the first born black hole in wide binaries), as lower mass black holes receive larger kicks than high-mass black holes in our model.
In addition to this, for tight binaries, the range of orbital periods that allow for chemically homogeneous evolution is narrow \citep[e.g.,][]{Mandel:2015qlu,Riley:2020btf}, and increased mass-loss can lead to the orbits of these binaries widening enough that the component stars can no longer evolve homogeneously.
We show the contribution to the intrinsic binary black hole merger rate from the chemically homogeneous evolution channel only in the bottom panel of Figure~\ref{fig:BBH_rate}, as well as in Figure~\ref{fig:fraction_CHE}. 
We see that both of the effects described above are much stronger for the population formed through chemically homogeneous evolution.
We also see a general trend in Figure~\ref{fig:BBH_rate} that models with larger values of $Z_0$ typically predict lower binary black hole merger rates, as the increased metallicity also corresponds to higher typical mass-loss rates.
This has a similar effect as changing the Wolf--Rayet mass-loss rates, as described above.
We do not see any clear trends with the efficiency of common envelope evolution, and this is likely due to the subdominant contribution of this population to the total merger rate in our model.
Similarly, we do not see any strong trends with $d$.

We have also overplotted the binary black hole merger rate inferred by \citet{LIGOScientific:2021psn} in Figure~\ref{fig:BBH_rate},
noting that these rates are inferred using a phenomenological mass distribution fit to the full population of binary black hole mergers. 
As discussed in the introduction, \citet{LIGOScientific:2021psn} estimate the intrinsic merger rate of binary black holes to be $\mathcal{R}_\mathrm{LVK} = 17$--$45$\,Gpc$^{-3}$\,yr$^{-1}$ at a redshift of $z = 0.2$\footnote{\citet{LIGOScientific:2021psn} quote the BBH merger rate at $z = 0.2$ as this is close to the redshift where most of the BBHs have been observed, and thus where the rate is best measured (see also \citealp{Roulet:2020wyq}).}.
This estimate is sensitive to the overall shape of the binary black hole mass distribution, particularly at low mass, where the intrinsic rate may be high but the observed rate may be low due to selection effects.
\citet{LIGOScientific:2021psn} infer the shape of the binary black hole mass distribution using a series of phenomenological models; our physical models directly predict both the intrinsic binary black hole merger rate and the shape of the mass distribution (as discussed further below).
We note that $\mathcal{R}_\mathrm{LVK}$ is the rate for the whole binary black hole population, integrated over all masses; since a significant fraction ($\sim 10\%$) of observed binary black holes have masses greater than can be produced through isolated binary evolution (see Section~\ref{sec:observational_sample}), we expect that the rate from isolated binary evolution of models matching observations should be less than that found by \citet{LIGOScientific:2021psn}. 
We discuss the constraints on our models implied by this comparison further in Section~\ref{subsec:compare_to_obs}.

We show the fraction of merging binary black holes formed through the chemically homogeneous evolution channel in Figure~\ref{fig:fraction_CHE}.
In the COMPAS fiducial model, around 70\% of observable binary black holes are formed through chemically homogeneous evolution \citep{Riley:2020btf}.
In our models, we find a large range in the fraction of merging binary black holes formed this way, from less than 20\% up to around 70\%.
Due to the strong gravitational-wave selection effects favouring the higher masses produced through the chemically homogeneous evolution channel (as discussed further in Section~\ref{subsec:BBH_mass_dist}), we find that more than 70\% of the observed population is formed through chemically homogeneous evolution in all of our models, and in some models up to 95\% of all observed binary black holes are formed through this channel.
We find that the largest fractions correspond to the models with the smallest values of $\alpha_\mathrm{CE}$, as these models have the smallest contribution from the classical common envelope channel \citep[][]{Belczynski:2016jno,Stevenson:2017tfq}.
We see that smaller fractions of observable binary black holes are formed through chemically homogeneous evolution in models with high values of $f_\mathrm{WR}$ (bottom panel of Figure~\ref{fig:fraction_CHE}).
This is because the enhanced mass loss causes many tight binaries to widen such that they can no longer merge within the age of the Universe.

\begin{figure}
    \centering
    \includegraphics[width=\columnwidth]{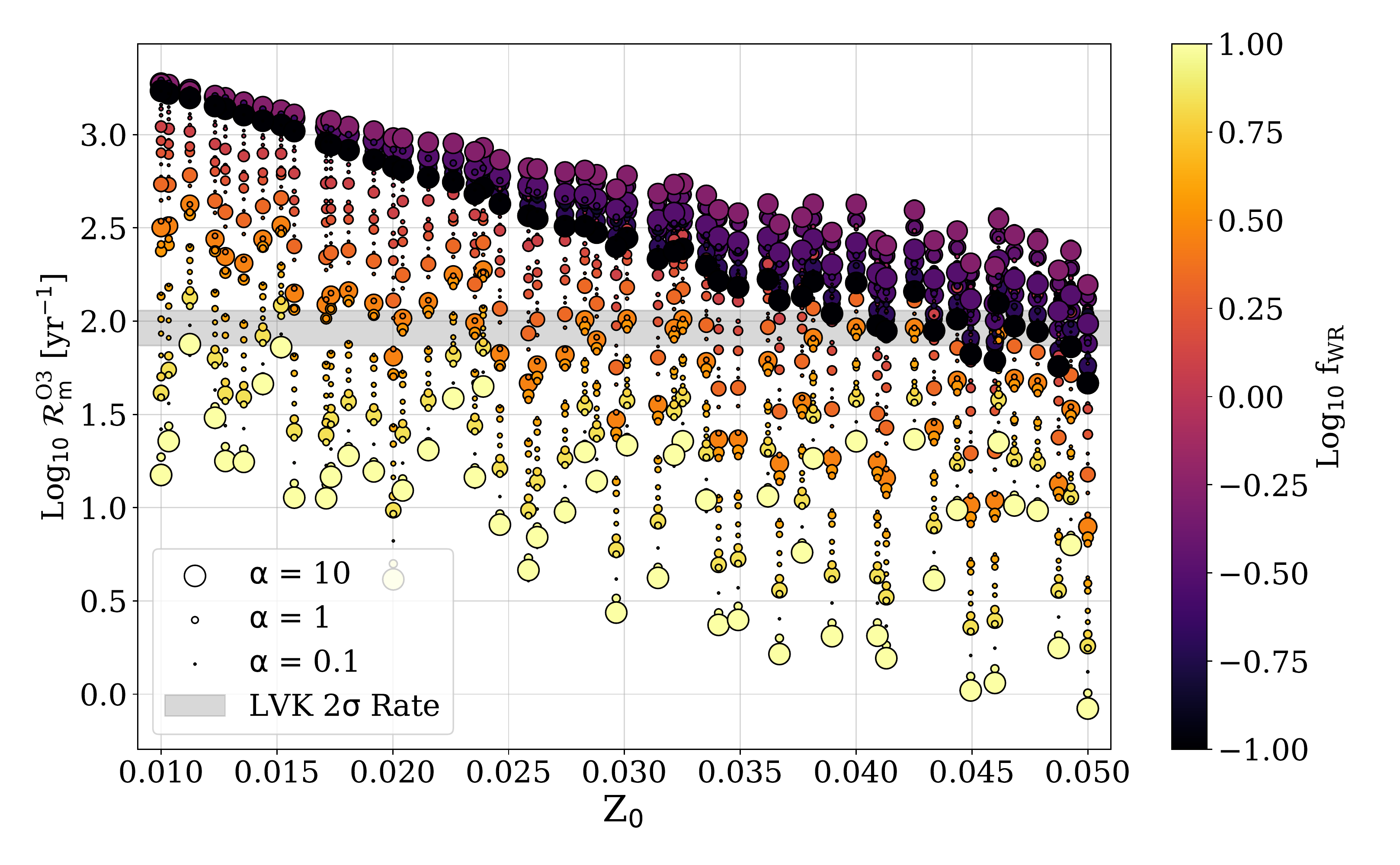}
    \includegraphics[width=\columnwidth]{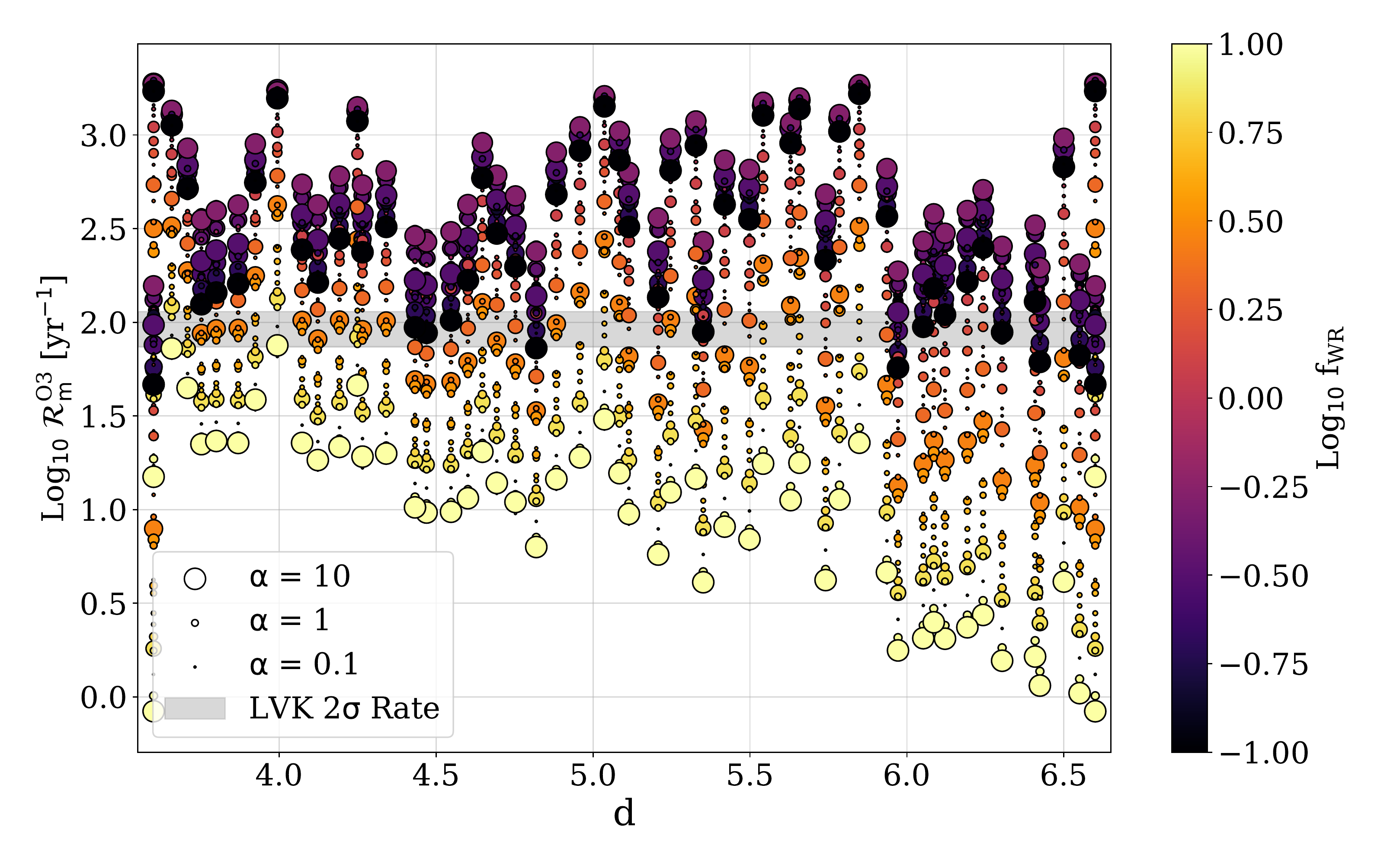}
    \caption{Predicted detection rate for O3 (calculated from Equation~\ref{eq:detection_rate}) for our set of models, shown as a function of model hyperparameters.
    The shaded gray region corresponds to the rate of binary black hole detections in O3 (as discussed in the text), accounting for $2\sigma$ Poisson uncertainties.
    The top panel shows the rate as a function of $Z_0$, $\alpha$ and $f_\mathrm{WR}$, whilst the bottom panel shows it as a function of $d$, $\alpha$ and $f_\mathrm{WR}$.
    }
    \label{fig:detection_rates}
\end{figure}

\begin{table*}
    \centering
    \begin{tabular}{c|c|c|c|c|c}
         Observing run & Catalogues & References & Duration ($T_\mathrm{obs}$) [d] & $N_\mathrm{BBH}$ \\
         \hline
         O1  & GWTC-1 & \citet{LIGOScientific:2016dsl,LIGOScientific:2018mvr} & 48 & 3 \\
         O2  & GWTC-1 & \citet{LIGOScientific:2018mvr} & 118 & 7 \\
         O3a & GWTC-2, GWTC-2.1 & \citet{Abbott:2021PhRvXGWTC-2,LIGOScientific:2021usb} & 149 (177) & 36 \\
         O3b & GWTC-3 & \citet{LIGOScientific:2021djp} & 126 (142) & 33 
    \end{tabular}
    \caption{Summary of the first three observing runs of Advanced LIGO and Virgo. We list the name of each observing run, the name and references for the  catalogue(s) from which we draw our observations, along with the duration of each observing run that at least two (one) interferometers were operational. 
    $N_\mathrm{BBH}$ denotes the number of binary black holes observed during each observing run.
    }
    \label{tab:observing_runs}
\end{table*}

The detection rates of binary black hole mergers predicted by our models (as calculated using  Equation~\ref{eq:detection_rate}) are shown in Figure~\ref{fig:detection_rates}, again as a function of model hyperparameters.
For a single Advanced LIGO detector operating at sensitivity comparable to that achieved during O3, our models predict 10--800 detections per year.
Assuming a duration for O3 of $T_\mathrm{obs} = 275$\,days \citep[][]{Abbott:2021PhRvXGWTC-2,LIGOScientific:2021djp}, the actual detection rate was around 100 per year.
The exact duration of data analysed varies between analysis pipelines \citep[][]{Abbott:2021PhRvXGWTC-2,LIGOScientific:2021djp}. 
However, we do not expect small changes in our assumed $T_\mathrm{obs}$ to qualitatively change our results.
We summarise the durations of each observing run and the number of detections made in Table~\ref{tab:observing_runs}.

\subsection{Binary black hole mass distribution}
\label{subsec:BBH_mass_dist}

In addition to impacting the overall rate of binary black hole mergers, we also expect that variations in our assumptions will lead to differences in the mass distribution of binary black holes.
Whilst it is difficult to summarise a distribution in a single number, here we opt to use the median observed chirp mass $\mathcal{M}_\mathrm{med}$ as a summary statistic in order to give some impression as to how the chirp mass distribution varies.

\begin{figure}
    \centering
    \includegraphics[width=\columnwidth]{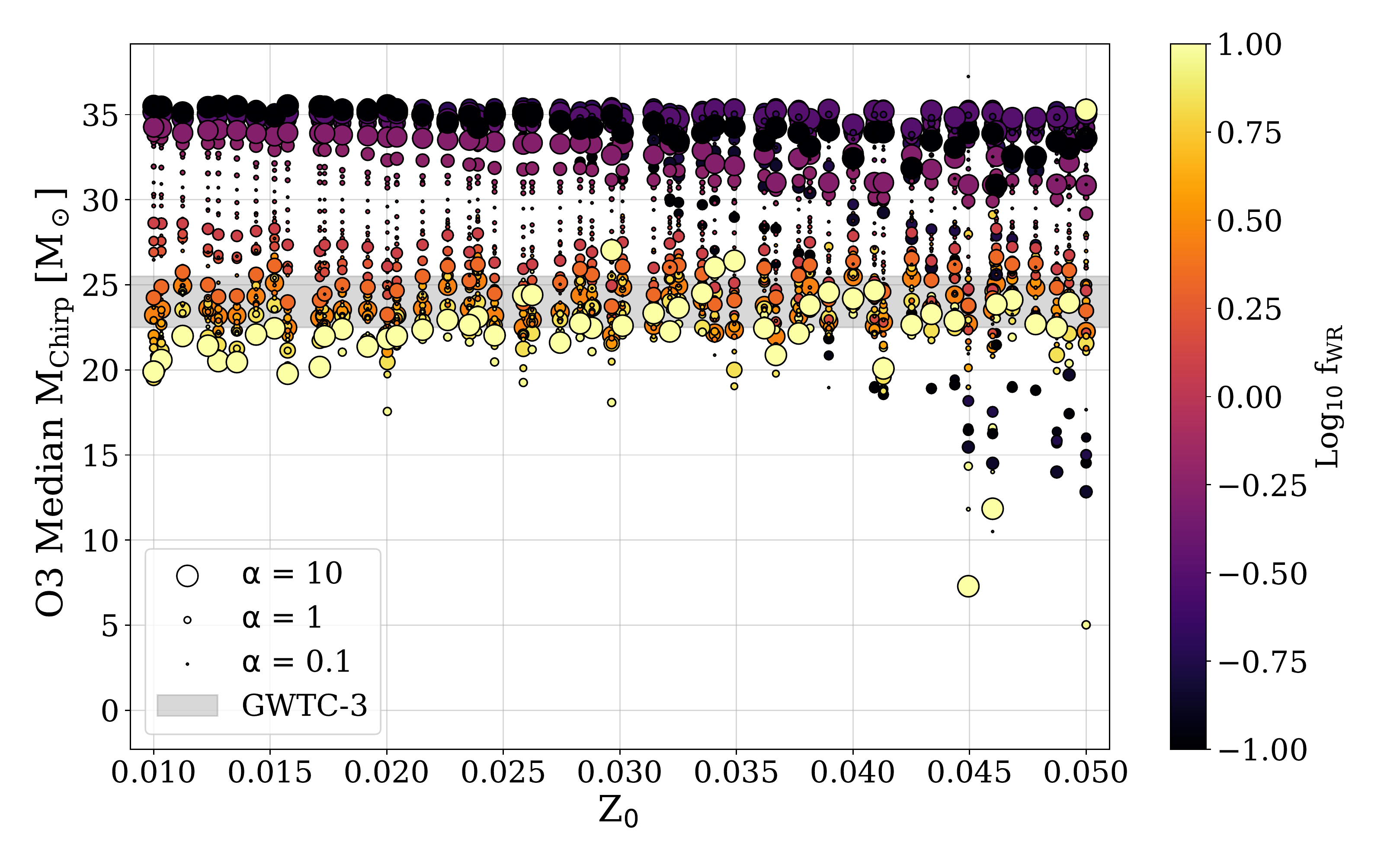}
    \includegraphics[width=\columnwidth]{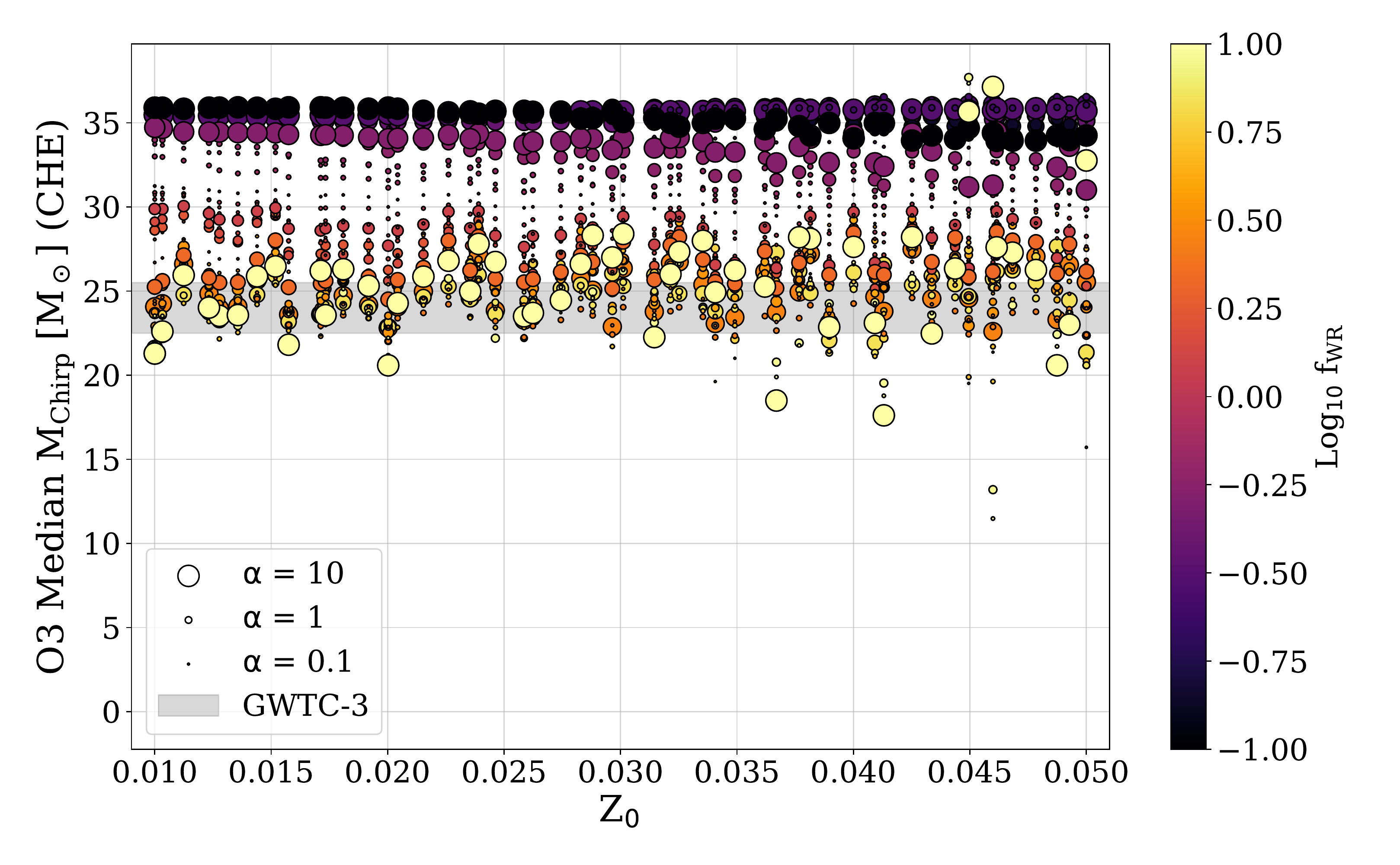}
    \includegraphics[width=\columnwidth]{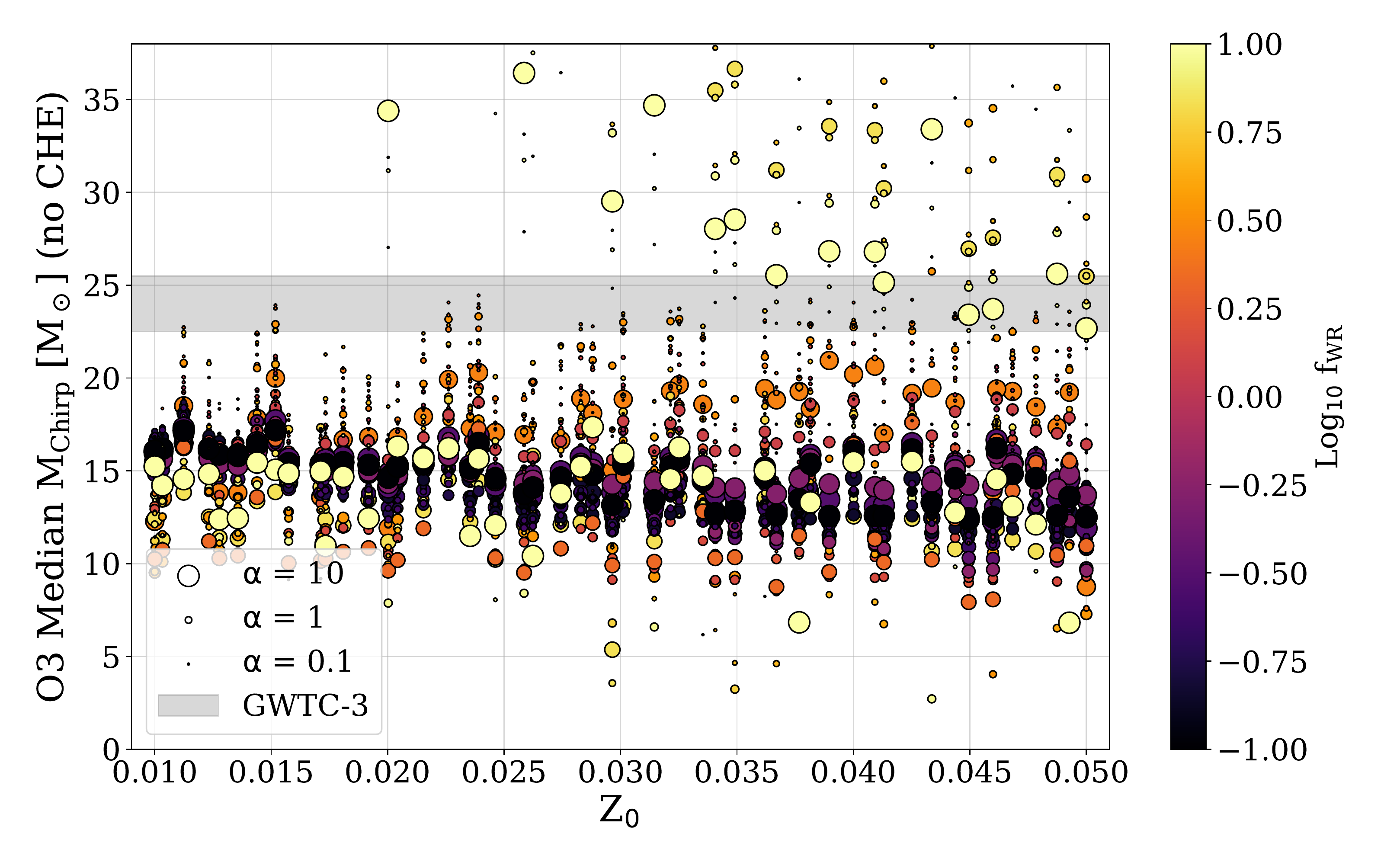}
    \caption{Median observed chirp mass of binary black holes predicted in our models as a function of the model hyperparameters. The top panel includes all merging binary black holes, whilst the middle panel shows only those formed through chemically homogeneous evolution and the bottom panel shows all formation channels except chemically homogeneous evolution. The shaded grey region indicates the median chirp mass of observed binary black holes \citep{LIGOScientific:2021djp}, as described in the text in Section~\ref{subsec:BBH_mass_dist}}
    \label{fig:median_chirp_mass_func_hyperparams}
\end{figure}

We show the median observed chirp mass predicted by each of our models as a function of the hyperparameters in Figure~\ref{fig:median_chirp_mass_func_hyperparams}.
We find that the distribution of chirp masses can vary quite significantly, with variations of more than $10$\,M$_\odot$ in predictions for the median chirp mass, ranging anywhere from $22$--$34$\,M$_\odot$.
We find that the median observed chirp mass has a strong dependence on the assumed mass-loss rates for Wolf--Rayet stars (as prescribed by our $f_\mathrm{WR}$ parameter).
In agreement with previous studies \citep[][]{Riley:2020btf}, we find that binary black holes formed through the chemically homogeneous evolution channel typically have higher average chirp masses ($24$--$34$\,M$_\odot$; middle panel of Figure~\ref{fig:median_chirp_mass_func_hyperparams}) compared to other isolated binary evolution formation channels ($10$--$23$\,M$_\odot$; bottom panel of Figure~\ref{fig:median_chirp_mass_func_hyperparams}). 
We find that neither $Z_0$ nor $d$ have a strong impact on the median observed chirp mass.
We also observe from the bottom panel of Figure~\ref{fig:median_chirp_mass_func_hyperparams} that models with the lowest values of $\alpha_\mathrm{CE}$ produce binary black holes with higher typical masses; we believe that this is because in these extreme models, most common envelope events result in mergers, and so this population then becomes dominated by the stable mass transfer channel, which typically produces more massive binary black holes than the common envelope channel \citep[][]{Neijssel:2019,Bavera:2020uch,vanSon:2021zpk}. 

In the bottom panel of Figure~\ref{fig:observed_chirp_mass_dist} we estimate the median of the observed chirp mass distribution. 
We determine that the median observed chirp mass is $24$--$25$\,M$_\odot$. 
We overlay this on Figure~\ref{fig:median_chirp_mass_func_hyperparams}.
Whilst some models are able to reproduce the average observed chirp mass when including binaries formed through chemically homogeneous evolution, most models do not predict enough massive binary black holes when excluding the chemically homogeneous evolution channel (bottom panel of Figure~\ref{fig:median_chirp_mass_func_hyperparams}).
We discuss the constraints implied for our models in Section~\ref{subsec:compare_to_obs}.

In our models, we find that binary black holes formed through chemically homogeneous evolution have chirp masses greater than 7\,M$_\odot$ \citep[][]{Riley:2020btf}.
\citet{duBuisson:2020asn} find merging binary black holes formed through chemically homogeneous evolution with chirp masses down to $15$\,M$_\odot$.
This would increase the fraction of the observed binary black holes consistent with forming through chemically homogeneous evolution (quoted in Section~\ref{sec:observational_sample}) to $\sim 70\%$.

\begin{figure}
    \centering
    \includegraphics[width=\columnwidth]{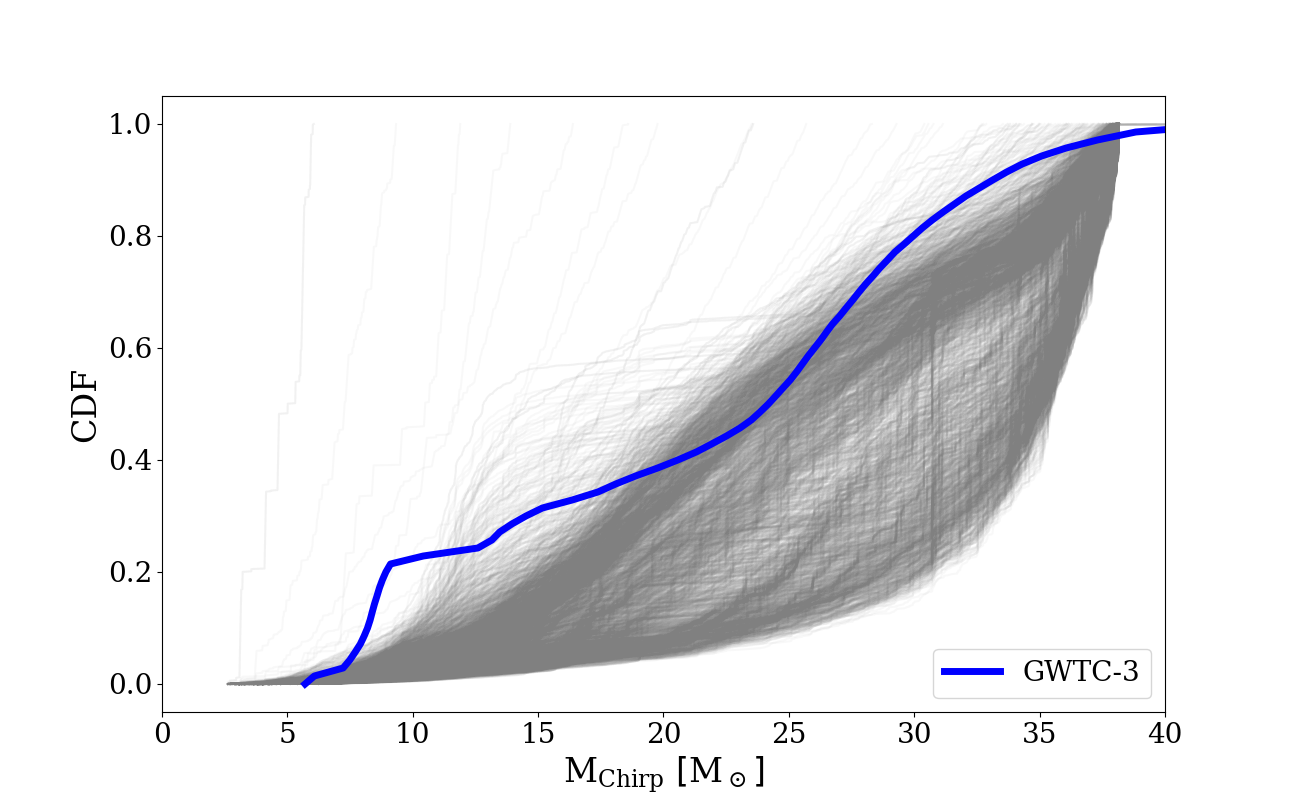}
    \caption{Cumulative observed chirp mass distributions predicted by all of our models are shown in gray, while the empirically determined chirp mass distribution (as shown in Figure~\ref{fig:observed_chirp_mass_dist}) is shown in blue. Our models typically predict too many high-mass binary black holes.}
    \label{fig:chirp_mass_dist_all_models}
\end{figure}

We show the chirp mass distributions predicted by our model in Figure~\ref{fig:chirp_mass_dist_all_models}. 
There is a large amount of variety in the chirp mass distribution of merging binary black holes predicted by COMPAS. 
However, we find that even among our broad range of models, none of these models produces a satisfactory match to the observed chirp mass distribution, with almost all models overpredicting the masses of binary black holes (our model distributions are shifted to the right compared to observations).
There are a few reasons why this may be the case.
Firstly, by construction we have limited the number of uncertain binary evolution assumptions we have explored, and other stages could also have a large impact on both the rate and mass distribution of merging binary black holes \citep[e.g.,][]{Broekgaarden:2021efa}.  
Secondly, with the exception of the events excluded in Section~\ref{sec:observational_sample}, we assume that all of the observed merging binary black holes formed through isolated binary evolution. 
Of course, this may not necessarily be true, and other channels may contribute to the binary black hole population in the mass range we consider as well.
\citet{Zevin:2020gbd} fit the observed binary black hole population using models of several different formation channels, and show that a mixture of different formation channels (including both isolated binary evolution and dynamical formation channels) can provide a good match to the overall properties of the observed binary black hole population.

\subsection{Comparison to observed rate and chirp mass distribution}
\label{subsec:compare_to_obs}

Since the focus of the paper is on exploring the predictions of the COMPAS population synthesis model as a function of the uncertain population parameters, a full comparison between the results presented here and the observed gravitational-wave population is deemed beyond the scope of this paper and left for future work (see also discussion in Section~\ref{sec:conclusions}).

However, it is of course still useful to do some simple comparisons in order to determine if the predictions from these models compare well with observations. 
We note that previous studies using COMPAS have made comparisons with gravitational-wave observations and found good agreement \citep[e.g.,][]{Stevenson:2017tfq,Neijssel:2019,Riley:2020btf,Broekgaarden:2021efa}.

\begin{figure*}
    \centering
    \includegraphics[width=\textwidth]{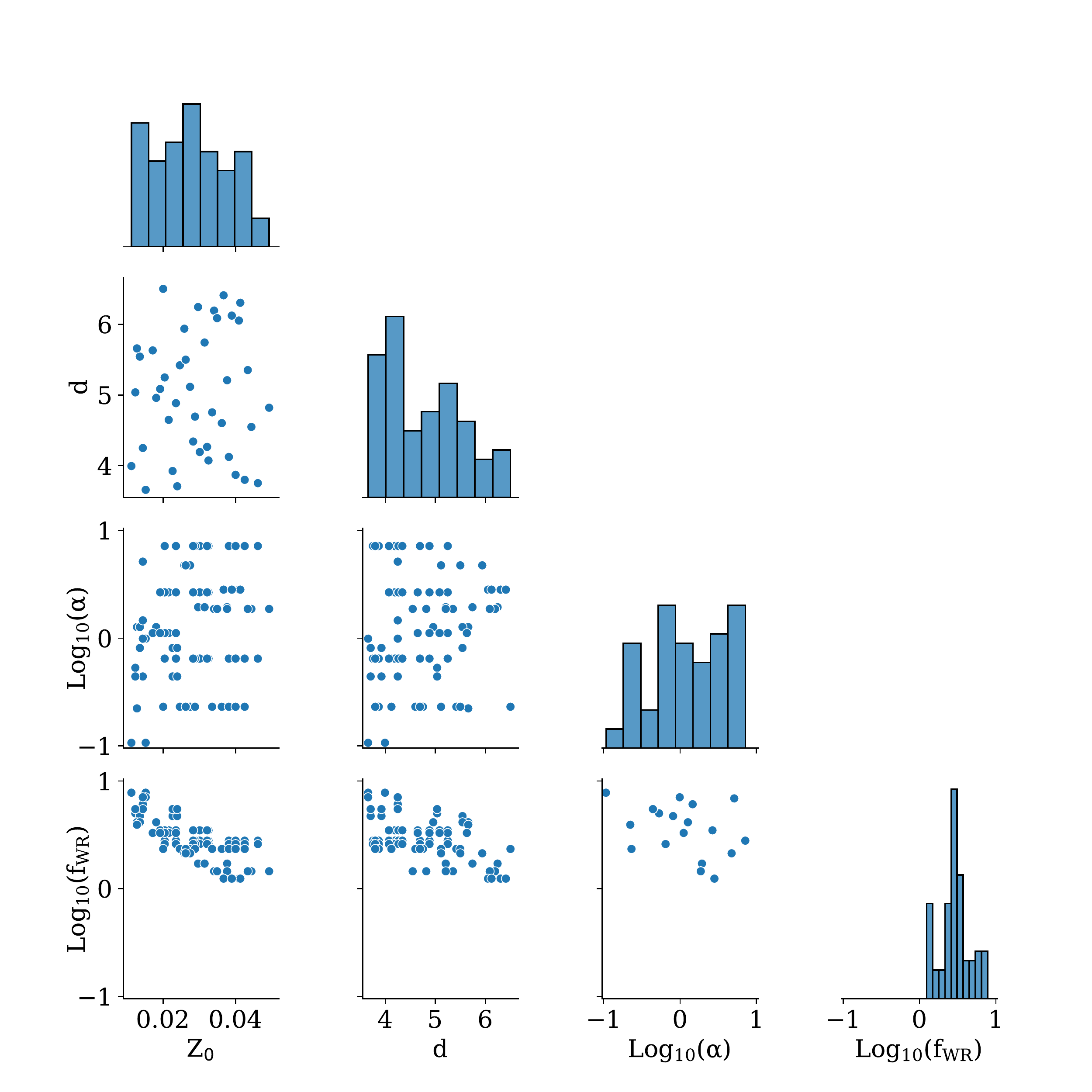}
    \caption{Matrix corner plot showing the one- and two-dimensional distributions of the hyperparameters for which our models match both the observed rate and average mass of binary black holes (for details, see Section~\ref{subsec:compare_to_obs}).
    }
    \label{fig:matching_hyperparameters}
\end{figure*}

We constrain our models by selecting only those models that simultaneously match both the observed binary black hole rate and the average binary black hole chirp mass, as shown in Figures~\ref{fig:detection_rates} and \ref{fig:median_chirp_mass_func_hyperparams}.
Specifically, for the observed binary black hole detection rate, we keep any model that predicts a rate similar to the observed rate in O3, within $2\sigma$ (90\%) Poisson uncertainties (as indicated by the shaded region in Figure~\ref{fig:detection_rates}). We opt to use the detection rate, rather than the merger rate, as the former is more constraining.
For constraining models based on their predicted mass distributions, we make a simple cut, keeping only those models that predict a median chirp mass in the range 22.5--25.5\,M$_\odot$, as observed.
We show all models that simultaneously match both of these constraints in Figure~\ref{fig:matching_hyperparameters}. 
Out of our 2,916 models, 145 models match the O3 detection rate, 583 models match the median chirp mass only and 79 models match both the median chirp mass and rate.

We find that only models with $f_\mathrm{WR} \gtrsim 1$ are capable of explaining both the observed rate of binary black hole mergers, and the mass distribution.
This predominantly removes some of the massive binary black holes formed through the CHE channel, both lowering the overall predicted merger rate and reducing the average mass (Figure~\ref{fig:fraction_CHE}).
As can be seen from Figure~\ref{fig:matching_hyperparameters}, this is somewhat degenerate with our choice of the typical metallicity of star formation ($Z_{0}$), since higher metallicities also lead to higher mass-loss rates \citep{Vink:2005zf}.
We have quantified the anti-correlation between $Z_{0}$ and $f_\mathrm{WR}$ by calculating both the Spearman and Pearson correlation coefficients, which are $-0.719$ and $-0.745$ respectively. Both of these report statistically significant anti-correlations, with $p$-values $p \ll 0.01$ in both cases.

A similar anti-correlation can be observed in Figure~\ref{fig:matching_hyperparameters} between $f_\mathrm{WR}$ and $d$, where large values of $d$ correspond to low star formation rates at high redshift (cf. Figure~\ref{fig:compare_SFR}), removing the contribution of low metallicity star formation at high redshift responsible for producing massive binary black holes (predominantly through chemically homogeneous evolution).
We again quantified the magnitude of this anti-correlation, finding significant ($p \ll 0.01$) Spearman and Pearson correlation coefficients of $-0.438$ and $-0.514$.
These degeneracies are examples of exactly the sorts of correlations that we set out to uncover, and can only be found by exploring the impact of multiple parameters simultaneously.

\begin{figure}
    \centering
    \includegraphics[width=\columnwidth]{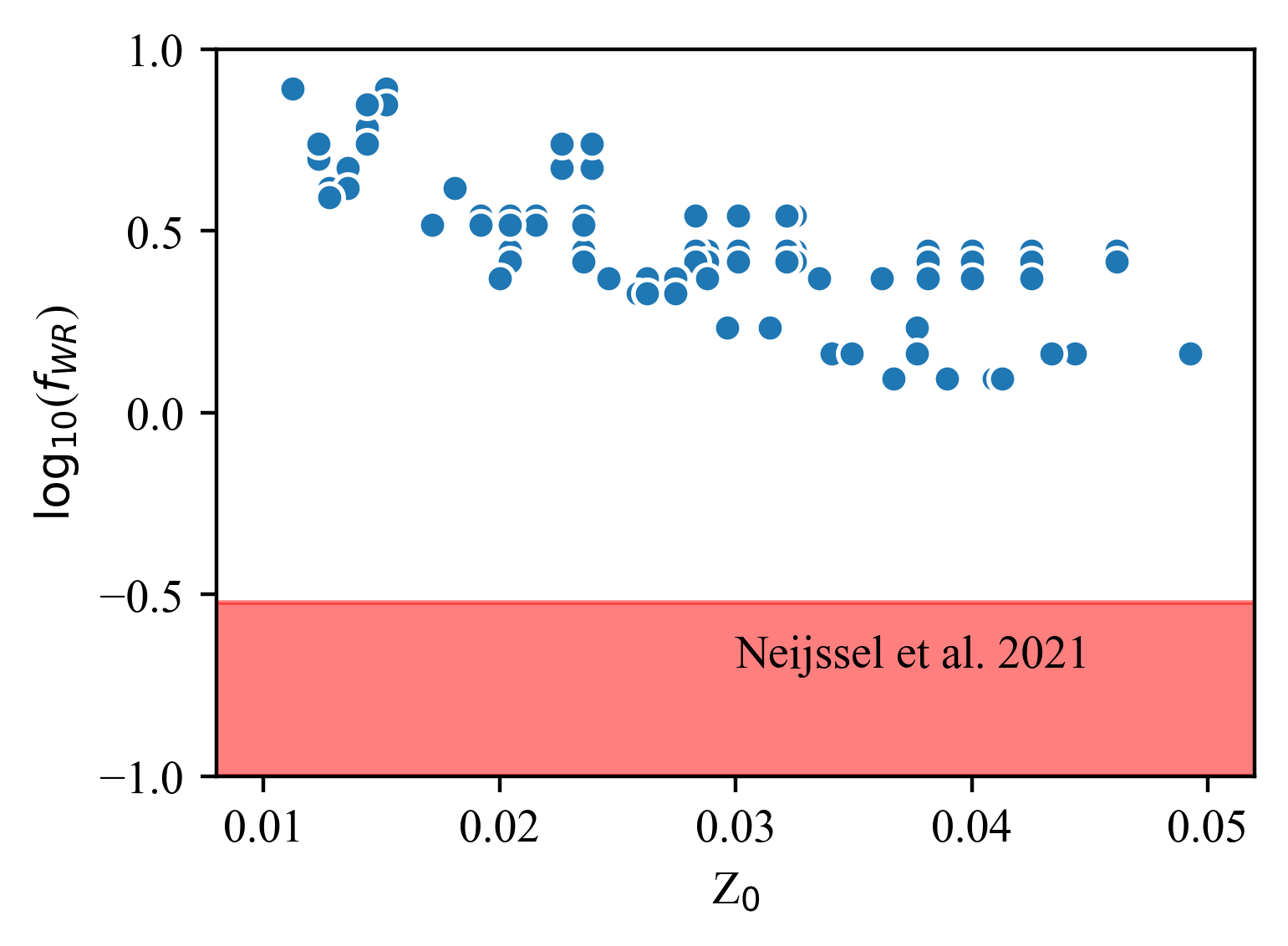}
    \caption{The correlation between the mass-loss rates of Wolf--Rayet stars ($f_\mathrm{WR}$) and the typical metallicity of star formation at redshift $z = 0$ ($Z_0$) for surviving models.
    The shaded red region shows the observational constraints placed on $f_\mathrm{WR}$ from the mass of the black hole in Cyg X-1 \citep[][]{Miller-Jones:2021plh,Neijssel:2021imj} showing the clear tension with our findings.
    }
    \label{fig:WR_bias}
\end{figure}

In Figure~\ref{fig:WR_bias} we zoom in on the correlation between $f_\mathrm{WR}$ and $Z_0$. 
We overplot the constraints that $f_\mathrm{WR} \lesssim 0.3$ from \citet{Neijssel:2021imj} based upon the mass of the black hole in Cyg X-1 \citep[][]{Miller-Jones:2021plh}. 
Our findings are in tension with both these recent observational developments, as well as recent theoretical work on Wolf--Rayet mass loss \citep[][]{Sander:2020MNRAS}. 
As we discuss below, this highlights potential biases that could arise when inferring binary evolution parameters due to limitations of our model (in particular, the modelling of chemically homogeneous evolution).

Most of our models with $\alpha_\mathrm{CE} < 1$ are ruled out (see Figure~\ref{fig:matching_hyperparameters}).
This is because binary black holes formed through the classical common envelope channel typically have the lowest masses out of the several isolated binary evolution subchannels modelled in COMPAS \citep[see e.g.,][]{vanSon:2021zpk}.
A population of low-mass binary black holes is required to match the observed mass distribution (Figure~\ref{fig:chirp_mass_dist_all_models}).
Other similar studies have also shown a preference for super-efficient common envelope evolution \citep[e.g.,][]{Santoliquido:2020axb,Wong:2020ise,Garcia:2021niy,Broekgaarden:2021hlu}.
\citet{Bouffanais:2020qds} performed a similar analysis to the one we have performed here. 
Using similar models of isolated binary evolution, \citet{Bouffanais:2020qds} vary the efficiency of mass transfer (assuming a single value for all mass transfer episodes with a non-degenerate accretor) and the efficiency of common envelope evolution. 
Only considering models with $\alpha_\mathrm{CE} > 1$, they find that their preferred models have $\alpha_\mathrm{CE} \sim 6$.

\begin{figure}
    \centering
    \includegraphics[width=\columnwidth]{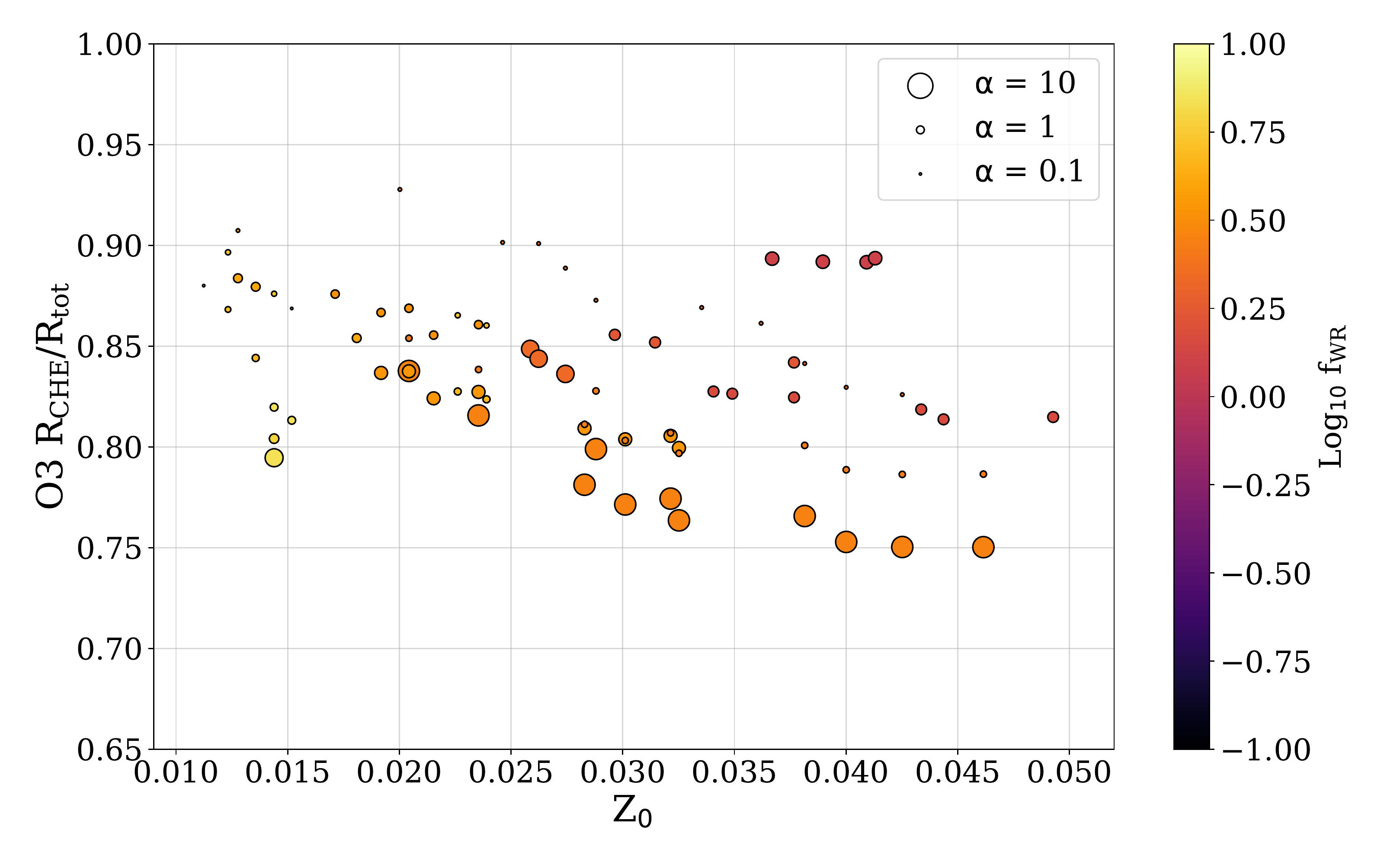}
    \includegraphics[width=\columnwidth]{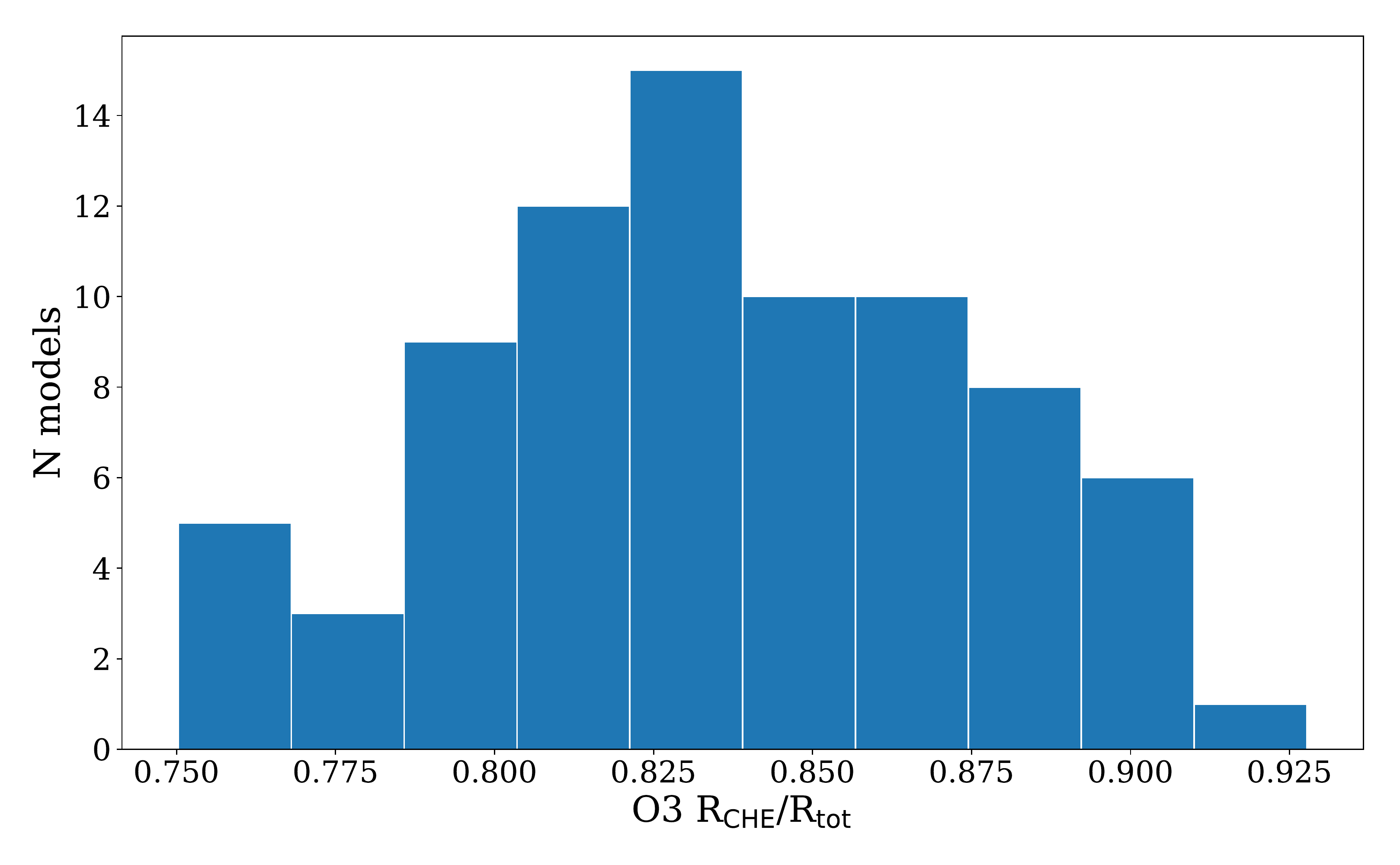}
    \caption{Fraction of merging binary black holes with chirp masses $< 40$\,M$_\odot$ formed through the chemically homogeneous evolution channel after constraining our models with observations. 
    The top panel shows the fraction as function of model hyperparameters, while the bottom panel shows a histogram of the fraction for all allowed models.}
    \label{fig:fraction_CHE_constrained}
\end{figure}

Our results are strongly sensitive to the inclusion of the chemically homogeneous evolution channel for forming binary black holes within COMPAS.
In Figure~\ref{fig:fraction_CHE_constrained} we show the fraction of binary black holes formed through chemically homogeneous evolution for the subset of models matching observations (as shown in Figure~\ref{fig:matching_hyperparameters}).
In models that simultaneously match both the observed binary black hole rate and median chirp mass, 75--90\% of binary black holes with chirp masses less than $40$\,M$_\odot$ are formed through chemically homogeneous evolution. 
At present, COMPAS is the only rapid population synthesis suite that self-consistently includes the chemically homogeneous evolution channel \citep{Riley:2020btf}, though see \citet{Ghodla:2022xnm} for details of a recent implementation in the BPASS code \citep[][]{Eldridge:2017PASA}.
Any constraints on the underlying physics of binary evolution obtained with rapid binary population synthesis codes that do not self consistently include chemically homogeneous evolution may be strongly biased at present.

\begin{figure}
    \centering
    \includegraphics[width=\columnwidth]{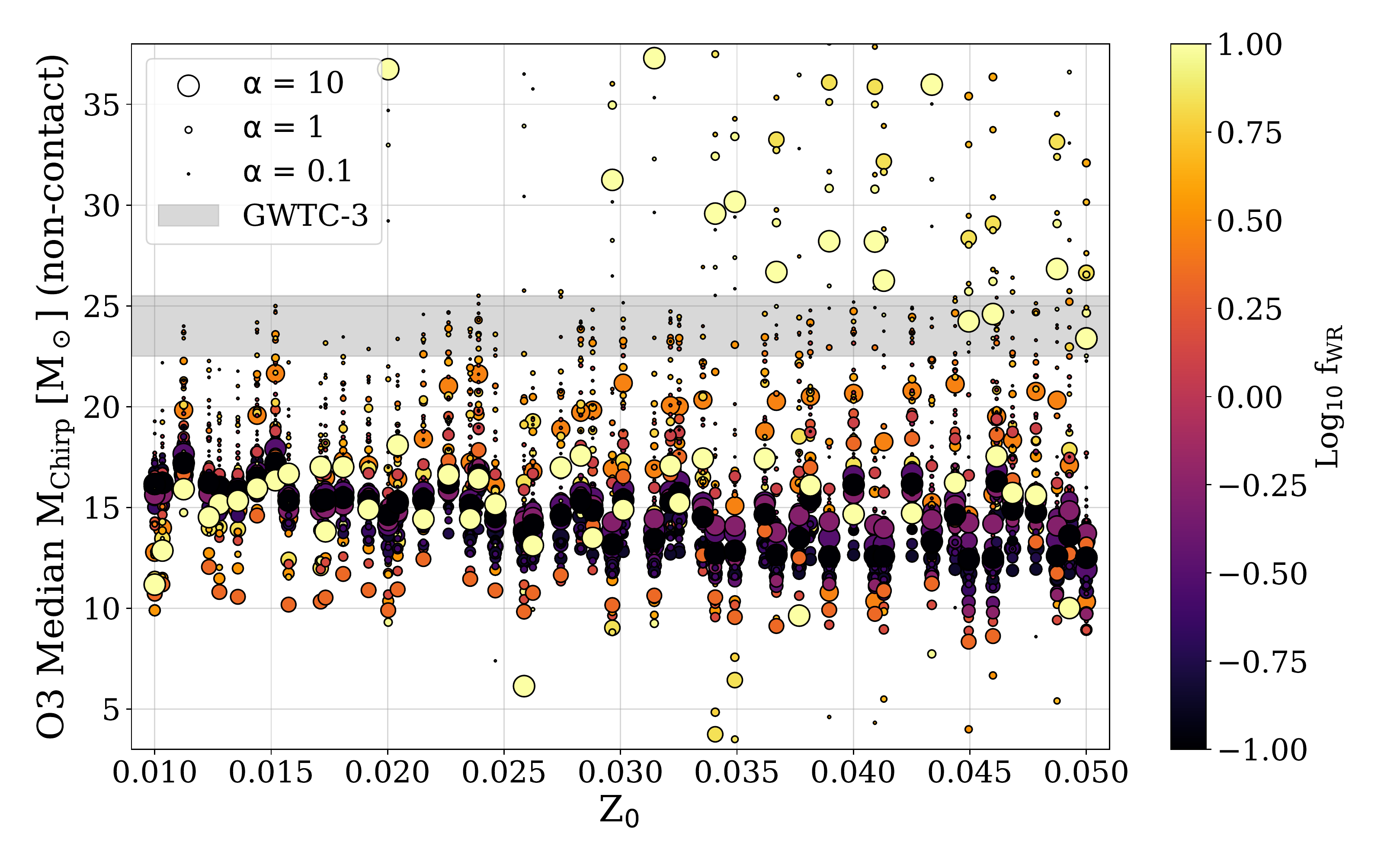}
    \includegraphics[width=\columnwidth]{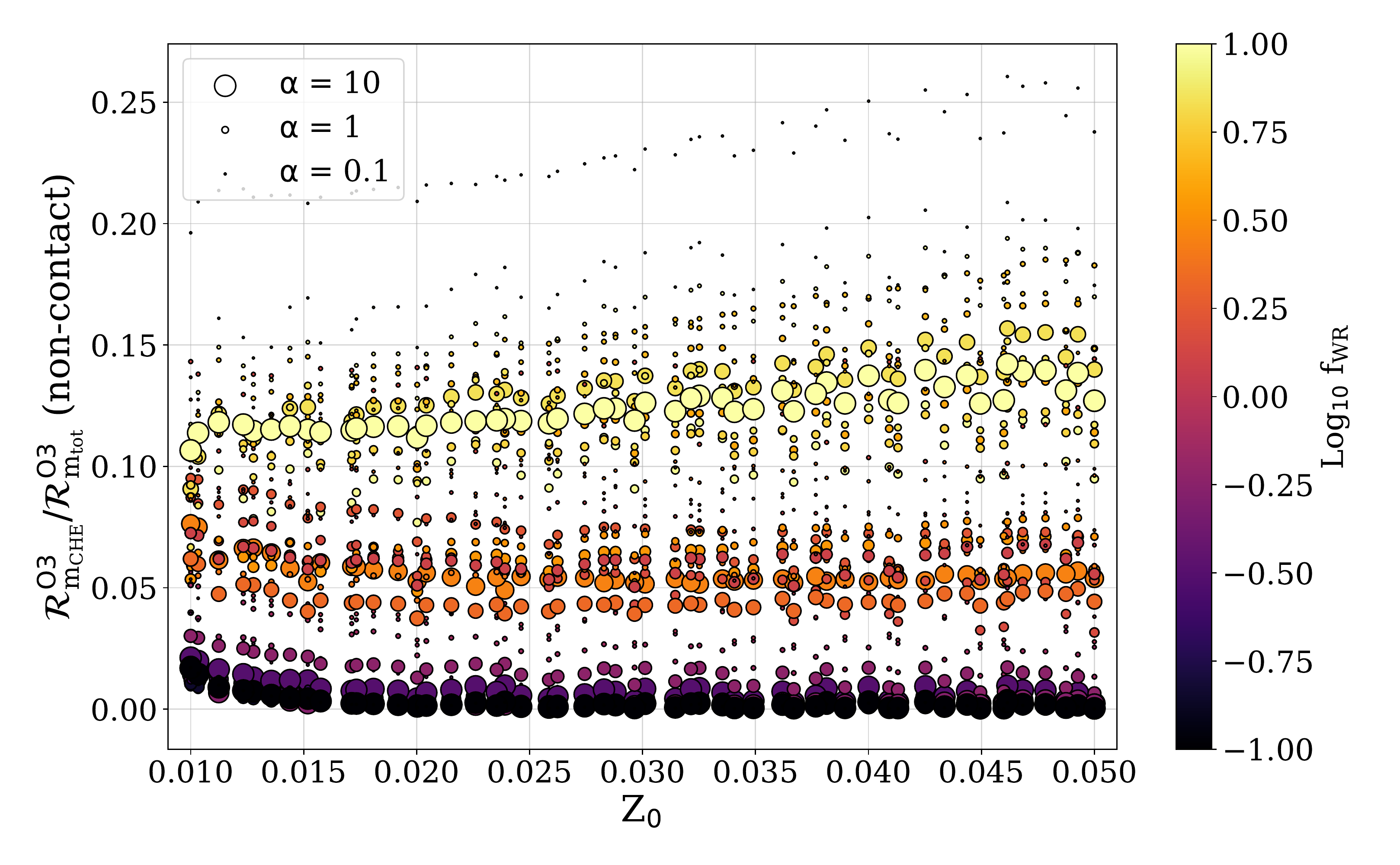}
    \caption{Top: The median chirp mass predicted to be observed in O3 as a function of model hyperparameters, discounting binaries which begin the main sequence in contact. 
    The plot is extremely similar to the distribution gained when no CHE binaries are included (cf. bottom panel of Figure~\ref{fig:median_chirp_mass_func_hyperparams}), since the majority of the binaries that undergo CHE in our simulations are actually over-contact binaries. 
    Bottom: The fraction of detected binary black holes formed through the CHE channel when we discount all binaries that start the main sequence as contact binaries. 
    This fraction is decreased dramatically (compared to Figure~\ref{fig:fraction_CHE}) when we exclude the contact binaries, as the majority of merging binary black holes formed through CHE begin their lives already in contact \citep[cf.][]{Marchant:2016wow}.}
    \label{fig:no_contact}
\end{figure}

We now turn to the question of why most of our models predict too many binary black holes with high chirp masses.
By default, the implementation of chemically homogeneous evolution in COMPAS \citep[][]{Riley:2020btf} allows for the stable evolution of binaries in contact at birth, so called massive overcontact binaries \citep[cf.][]{Marchant:2016wow}. 
However, the physics of contact binaries is not well understood \citep[][]{Abdul-Masih:2022arXiv}, and these overcontact binaries could be responsible for the discrepancy between our model predictions and observations.
As an alternate model, we investigated how our predictions would change if we excluded binaries that are in contact at birth. 
Figure~\ref{fig:no_contact} shows the median chirp mass and the fraction of binary black holes formed through CHE detected in O3 neglecting over-contact binaries.
We find that the predictions are similar to when completely excluding all binaries formed through chemically homogeneous evolution, as most of these binaries begin their evolution as over-contact binaries.
When excluding over-contact binaries from the population, these models typically underpredict the median binary black hole chirp mass compared to the observed value, as $\lesssim 20\%$ of binary black holes are predicted to be formed through chemically homogeneous evolution due to the small parameter space available avoiding Roche-lobe overflow \citep[][]{Marchant:2016wow,Riley:2020btf}.

\begin{figure}
    \centering
    \includegraphics[width=\columnwidth]{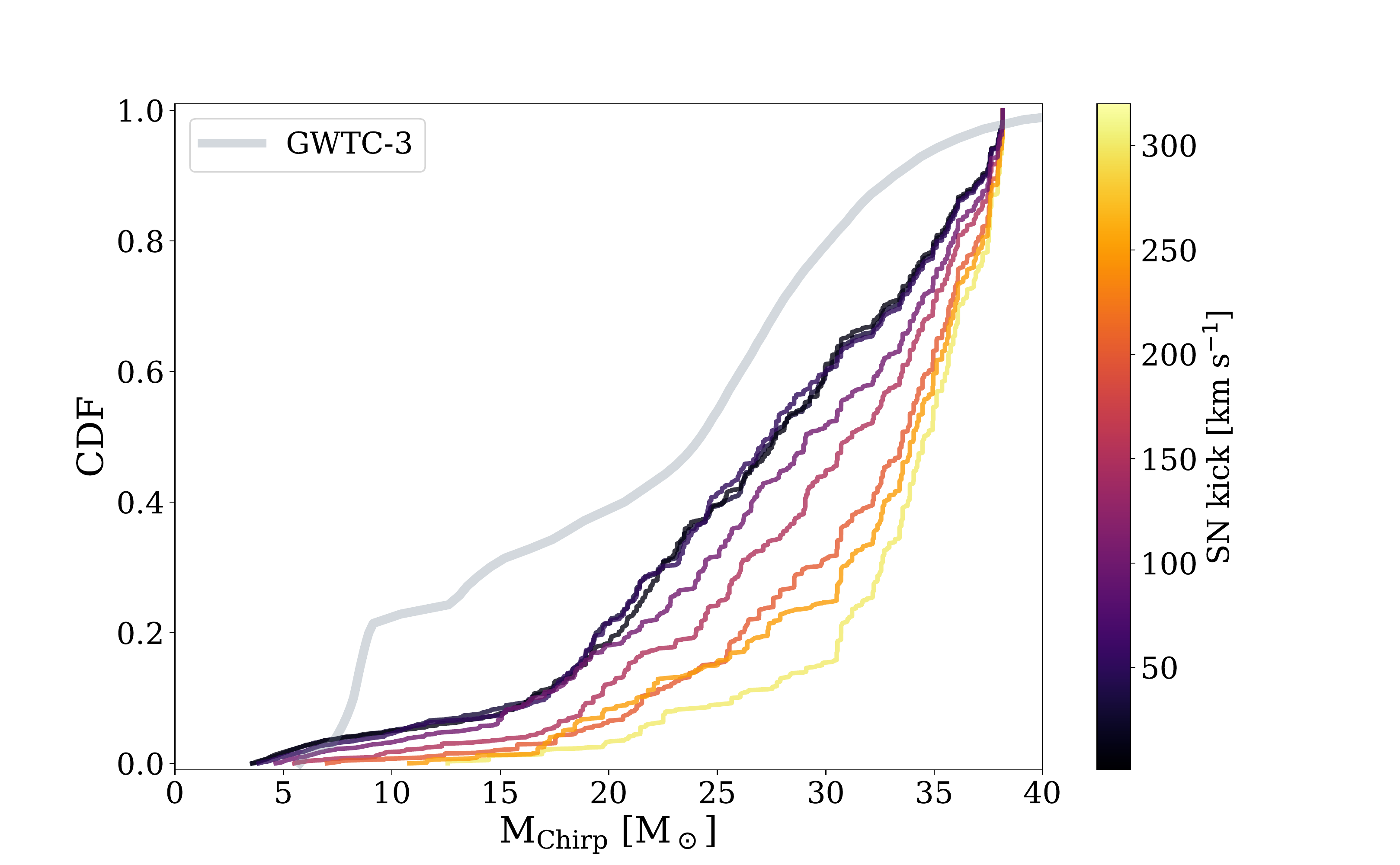}
    \caption{Cumulative density functions of the distribution of binary black hole chirp masses computed for 
    eight populations with different assumptions about the typical supernova kick magnitudes black holes receive at formation. 
    The other population parameters were kept at their default values in COMPAS \citep[][]{COMPAS:2021methodsPaper}. 
    For comparison, the empirical mass distribution obtained following GWTC-3 shown in black \citep[][]{LIGOScientific:2021djp}. 
    While smaller kicks associated with black hole formation shift the distribution towards the observed distribution, none of the kicks tested can produce predictions which satisfactorily match the observations.}
    \label{fig:SN_kicks}
\end{figure}

Another element of binary evolution that we have neglected in this paper so far is kicks associated with black hole formation. 
In our default model used elsewhere in this paper, the kicks of black holes are significantly reduced compared to those of neutron stars through fallback \citep[][]{Fryer:2012ApJ}, resulting in many of the most massive black holes receiving no kick at formation.
Black hole kicks would act to disrupt more binaries, reducing our overall predicted merger rates, and could alter the mass distribution of binary black holes \citep[e.g.,][]{Wysocki:2017isg}.
We have performed some test simulations to investigate whether allowing for large kicks associated with black hole formation could explain the difference between our models and the observed population. 
We calculated some test models where we include the impact of large kicks associated with black hole formation, drawn from a Maxwellian distribution and not reducing the kick magnitude due to fallback.
Figure~\ref{fig:SN_kicks} shows the results of varying the typical kick magnitude on the chirp mass distribution of the models, keeping all other parameters fixed to the COMPAS defaults \citep[][]{COMPAS:2021methodsPaper}.
For all models the mass distribution is still peaked at higher masses than the observations, with the lowest kicks (which are close to our default assumptions) being closest to observations. 
This is likely because our model predictions are dominated by high-mass binary black holes formed through chemically homogeneous evolution, which would require extremely large kicks to be disrupted.
We conclude that moderate to high black hole kicks cannot explain the discrepancy between our models and observations.
Regardless, such high black hole kicks are strongly disfavoured by independent observations \citep[e.g., see discussion in][]{Callister:2020vyz}.

\section{Discussion and conclusion}
\label{sec:conclusions}

In this paper we have investigated the correlated impact of multiple uncertainties in the evolution of massive binary stars across cosmic time on the population of merging binary black holes observable by current ground-based gravitational-wave observatories.

We made use of the rapid binary population synthesis suite COMPAS \citep{Stevenson:2017tfq,Vigna-Gomez:2018dza,COMPAS:2021methodsPaper}.
We focused on a few specific examples of uncertain evolutionary stages in massive binary evolution. 
Specifically, we investigated: 
\begin{itemize}
    \item the efficiency of common envelope evolution ($\alpha_\mathrm{CE}$)
    \item the mass loss rates of helium-rich Wolf--Rayet stars ($f_\mathrm{WR}$)
    \item the cosmic star formation rate at high redshift ($d$)
    \item the average gas-phase metallicity of star forming material ($Z_0$)
\end{itemize}
We simulated a large number ($N = 2916$) of binary populations, where multiple parameters were allowed to vary from their default values (cf. Figure~\ref{fig:parameter_space_cartoon}).
Our goal was to fully explore the (hyper)parameter space of this population synthesis model, and identify correlations or degeneracies between multiple population hyperparameters.
We identified a correlation between the impact of the mass-loss rates for Wolf--Rayet stars ($f_\mathrm{WR}$) and the average metallicity of star formation at redshift 0 ($Z_0$) on the properties of merging binary black holes, as can be seen clearly in Figure~\ref{fig:BBH_rate} and the bottom left panel of Figure~\ref{fig:matching_hyperparameters}. 
This correlation arises as increasing both parameters leads to increase in the amount of mass lost through stellar winds.

Our primary conclusion is that a large amount of variation is possible within models of isolated binary evolution given present uncertainties (cf. Figure~\ref{fig:BBH_rate} and Figure~\ref{fig:chirp_mass_dist_all_models}).
Whilst many of our models can produce binary black hole merger rates in agreement with observations (cf. Figure~\ref{fig:BBH_rate}), none of the models we consider provide a good match to the observed binary black hole mass distribution (Figure~\ref{fig:chirp_mass_dist_all_models}).
The models that are closest to the observations require enhanced (high) mass-loss rates for helium rich Wolf--Rayet stars ($f_\mathrm{WR} > 1$, see Figure~\ref{fig:matching_hyperparameters}).
This is because high mass-loss rates reduce both the mass of black holes, and the overall binary black hole merger rate, particularly of those formed through chemically homogeneous evolution (Figure~\ref{fig:fraction_CHE}).
Such high mass-loss rates are in tension with recent theoretical and observational developments regarding the winds of Wolf--Rayet stars \citep[e.g.,][]{Sander:2020MNRAS,Neijssel:2021imj}.

There are a number of possible reasons that our models tend to produce binary black holes that are too massive in comparison to observations.
For example, we have restricted our analysis to the four hyperparameters described above, both to limit the computational cost of the model exploration (as discussed in Section~\ref{subsec:model_exploration}) and to aid in interpreting the results.
However, there are a number of additional uncertainties in massive binary evolution which are also important in predictions for binary black holes \citep[][]{Broekgaarden:2021efa}. 

Future work should also expand the analysis presented here to other population hyperparameters, corresponding to other uncertain stages of massive binary evolution beyond those varied in this study.
This could include mass-loss rates during other stages of binary evolution \citep{Barrett:2017fcw}, the efficiency and stability of mass transfer \citep{Kruckow:2018slo,Broekgaarden:2021efa,Bouffanais:2020qds,Bavera:2021evk}, and supernova kicks imparted to neutron stars and black holes at birth \citep[e.g.,][]{Zevin:2017evb,Wysocki:2017isg,Callister:2020vyz,Stevenson:2022hmi}.
Similarly, other hyperparameters  governing the metallicity specific star formation rate (including the particular parameterisation) should be varied in order to fully explore its impact on the population of merging binary black holes \citep[see e.g.,][]{Chruslinska:2022arXiv,vanSon:2022ylf}.
As mentioned earlier, some parameters may be more important for some populations than others \citep[cf.][]{Broekgaarden:2021efa}.
Another uncertainty that we have not accounted for in the present analysis concerns the evolution of massive stars. 
Even detailed stellar models can disagree wildly on the properties of massive stars, which can have a dramatic impact on predictions for the formation of merging binary black holes \citep[][]{Marchant:2021hiv,Klencki:2020kxd,Gallegos-Garcia:2021hti,Agrawal:2021arXiv}.
Work is underway to allow these uncertainties to be incorporated into rapid population synthesis codes \citep[][]{Kruckow:2018slo,Spera:2019MNRAS,Agrawal:2020MNRASMETISSE,Fragos:2022}.
We leave exploring the impact of these parameters to future work.

We have focused our model exploration primarily on predictions for the mass distribution and merger rate of binary black holes.
Another key gravitational-wave observable is the spin of black holes \citep[e.g.,][]{Wysocki:2017isg,Gerosa:2018wbw,Belczynski:2017gds,Bavera:2019,Bavera:2020uch,Broekgaarden:2022nst}.
Recent binary evolution models typically predict that the first born black hole is born with negligible spin due to the majority of the stars' angular momentum being efficiently transported from the stellar core to the envelope  \citep[][]{Spruit:2001tz,Fuller:2019MNRAS} and then removed through stellar winds and binary mass transfer  \citep[][]{Fuller:2019sxi,Qin:2018vaa}.
The progenitor of the second born black hole may be tidally spun up if the orbital period of the binary is short enough, leading to some fraction of second born black holes having rapid rotation \citep[][]{Qin:2018vaa,2018MNRAS.473.4174Z,Bavera:2019,Bavera:2021evk}.
Mass ratio reversal occurring through mass transfer may allow for the more massive black hole to be born second in some fraction of binary black holes, allowing for the possibility of them to be rapidly rotating \citep{Zevin:2022wrw,Broekgaarden:2022nst}.
Binary black holes formed through chemically homogeneous evolution may have large aligned spins \citep{Marchant:2016wow}. 
It is likely that some of the hyperparameters we have varied here, such as the mass-loss rates of Wolf--Rayet stars, will have an observable impact on the spin distribution of observable binary black holes.

Our population synthesis model assumes a universal efficiency for the common envelope phase of massive binary evolution.
The common envelope efficiency parameter may not be universal, and may for example depend on the properties of the binary (such as the mass ratio) at the time the common envelope phase occurs \citep{DeMarco:2011MNRAS,Davis:2012MNRAS}.
Since binary black holes may form from  progenitors with similar properties (in terms of significant mass ratios and wide orbits) it may be that this approximation is not too bad in this case.
However, it is likely that, even if gravitational-wave observations do provide a precise measurement of $\alpha_\mathrm{CE}$, that should be considered as only applying to massive binary black hole progenitors, and a different value of $\alpha_\mathrm{CE}$ may be applicable for other populations \citep[for example,][find that $\alpha_\mathrm{CE} \sim 0.2$--$0.3$ provides a good match to a sample of low-mass post common envelope binaries consisting of white dwarfs and main sequence stars]{Zorotovic:2010A&A}.
Futhermore, the energy formalism \citep{Webbink:1984ApJ} may not be the correct description of the common envelope for massive stars \citep[e.g.,][]{Nelemans:2000A&A}.
We argue that even in this case, our results may still be interpretable. 
For example, some of our more extreme models (with $\alpha_\mathrm{CE} \ll 1$) result in a dramatic reduction in the formation rate of binary black holes through common envelope evolution as many binaries that would otherwise form that way end up merging during the common envelope phase (cf. Figure~\ref{fig:fraction_CHE}). 
These models may therefore still be relevant if binary black hole formation through common envelope evolution is rare, even if that rarity is due to another reason, as discussed by \citet{Klencki:2020kxd} and \citet{Marchant:2021hiv}. 

In this paper we focused on binary black holes as the these have the largest number of observations to compare against so far. 
Recent studies have shown that different populations of double compact objects are sensitive to different binary evolution physics \citep[][]{Broekgaarden:2021iew,Broekgaarden:2021efa}.
For example, \citet{Broekgaarden:2021efa} find that predictions for binary neutron stars are not sensitive to uncertainties in the cosmic star formation history, since in these models, the formation of neutron star binaries is much less sensitive to metallicity than black hole formation. 
Similar results have been found by others including \citet{Tang:2019qhn}.
In the future, we will need to expand our analysis to include both binary neutron star and neutron star-black hole binaries.
We note that previous work has shown that predictions from COMPAS are broadly in agreement with the observed rates and properties of binary neutron star and neutron star-black hole binaries for a range of model choices \citep[][]{Vigna-Gomez:2018dza,Chattopadhyay:DNS2019,Chattopadhyay:2020lff,Broekgaarden:2021efa,Broekgaarden:2021hlu,Broekgaarden:2021iew,Chattopadhyay:2022MNRAS}.

Several observed gravitational-wave events have properties that are difficult to explain through isolated binary evolution (e.g., GW190521 and GW190814), though several groups of authors have proposed counterarguments \citep{Zevin:2020gma,Antoniadis:2021dhe,Costa:2020MNRAS,Belczynski:2020bca}.
We have assumed that any binary black hole with a chirp mass greater than $40$\,M$_\odot$ cannot be formed through isolated binary evolution, consistent with our models \citep[though see][for counter arguments]{Costa:2020MNRAS,Belczynski:2020bca,Liu:2020lmi,Kinugawa:2020xws}.
Under this assumption, we argue that at least $10\%$ of observed binary black hole mergers must have a formation channel other than classical isolated binary evolution of massive, metal poor population I/II stars (Figure~\ref{fig:frac_too_massive}).
Leading candidates include dynamical formation in dense stellar environments such as young star clusters \citep[][]{DiCarlo:2019fcq}, old globular clusters \citep[][]{Rodriguez:2017pec} or the discs around active galactic nuclei \citep[][]{Yang:2019cbr}.
We similarly argued that the 40\% of binary black holes with chirp masses less than $20$\,M$_\odot$ cannot have formed through chemically homogeneous evolution \citep[][]{Mandel:2015qlu,deMink:2016vkw}, placing an upper limit of around $50$\% on the fraction of detected binary black holes that can have formed through that channel.
These inferences qualitatively agree with the findings of \citet{Zevin:2020gbd} (see their Figure 4) who used a mixture model consisting of population synthesis model predictions for several different formation scenarios to constrain the fraction of binary black holes formed through each channel. 
Whilst our approach is somewhat different, we find this overall agreement reassuring.

Regardless of the specific subchannel, we have assumed that all binary black holes with chirp masses less than $40$\,M$_\odot$ are formed through isolated binary evolution. 
However, it is of course possible (and even likely) that there are additional contributions from these other channels, though determining the formation channel for any given event is extremely difficult. 
Attempting to constrain binary evolution parameters whilst including events formed through other channels will inevitably lead to biases in the estimates of the population parameters (and thus of our understanding of binary evolution). 
One approach that can be employed to mitigate this issue is to combine models of multiple formation scenarios, and include the branching ratios between these scenarios as an additional parameter to be fit \citep[e.g.,][]{Zevin:2017evb,Zevin:2020gbd,Stevenson:2017dlk,Wong:2020ise,Bouffanais:2019ApJ,Bouffanais:2021wcr}.
Another way to mitigate these biases is to complement modelled analyses with model independent analyses that search for subpopulations of gravitational-wave sources with similar properties 
\citep[][]{Mandel:2016prl,Powell:2019nmw}.
Another possibility is to infer the population hyperparameters that best reproduce each individual event; if extreme assumptions are required to explain a particular event, it again might indicate a different formation scenario \citep[][]{Wong:2022arXiv}.

This work is intended to form the first stage of our analysis of gravitational-wave observations.
One of our eventual goals is to be able to constrain these parameters (and thus massive binary evolution) by comparing our models to observations \citep[][]{Barrett:2017fcw}.
Even though population synthesis codes such as COMPAS are extremely fast, capable of simulating a population of $10^{6}$ binaries in $\sim 12$\,hrs, it is still infeasible to directly use COMPAS to fully explore the parameter space, where potentially millions (or even more) of likelihood evaluations would be required.
One can think of our exploration here as a coarse, manual exploration of a restricted parameter space, rather than an automated exploration using a stochastic sampling technique such as Markov-Chain Monte Carlo or Nested Sampling \citep{SkillingNestedSampling}.
There are several approaches one could take in order to fully explore the (hyper)parameter space with these approaches.
One option could be to construct an emulator which can interpolate the predictions of a limited set of population synthesis models. 
Some early work exploring interpolation of population synthesis models was made by \citet{Barrett:2016edh} and \citet{Taylor:2018iat}.
This approach typically utilises methods such as machine learning (e.g., deep flow, random forest regression or neural networks) or Gaussian process regression \citep{Barrett:2016edh,Taylor:2018iat,Lin:2021evl,Wong:2019uni,Wong:2020ise}.
These approximate, emulated models are then fast enough to evaluate to be used in a likelihood evaluation in a stochastic sampler, whilst also allowing for the model to be evaluated at any arbitrary set of hyperparameters. 
Another option (which is less explored in this context) would be to perform inference with a sparse set of models, and then interpolate the likelihood \citep[e.g.,][]{Smith:2013zya,Abbott:2016PhRvD}. 

As with all population models, detailed binary evolution models of chemically homogeneous evolution have several important uncertainties.
One of the primary sources of uncertainty is the mass-loss rates, as already discussed in Section~\ref{subsec:winds}.
In addition to the mass-loss rates of helium rich stars, the mass-loss rates of hydrogen rich, chemically homogeneously evolving main-sequence stars are likely underestimated in COMPAS, as the fits in COMPAS assume that these stars have the same luminosity as a `normal' main-sequence star of the same mass and age \citep[][]{Riley:2020btf}.
However, a chemically homogeneously evolving star is expected to be more luminous than a conventionally evolving star of the same mass, which would lead to increased mass-loss rates on the main-sequence. 
This could potentially cause some binaries to widen sufficiently to exit the parameter space for chemically homogeneous evolution. 
Beyond the uncertainties in the mass-loss rates, the main uncertainty in models of chemically homogeneous evolution involves the treatment of rotationally enhanced mixing in one-dimensional stellar evolution codes, and the range of masses, rotation rates and orbital periods that allow for chemically homogeneous evolution \citep[see][for deeper discussion]{Mandel:2015qlu}.
Chemically homogeneous evolution is predominantly a theoretical phenomena at present, although a handful of observations over recent years have highlighted systems which can potentially (only) be explained through this channel \citep{Martins:2013A&A,Almeida:2015ApJ,Abdul-Masih:2021A&A}. 
Attempts to constrain the chemically homogeneous evolution pathway through observations of stars are difficult due to the strong preference for low metallicity environments.
Whilst we found a preference for models with high Wolf--Rayet mass-loss rates (and correspondingly lower binary black hole merger rates), an alternative interpretation could be that our model allows too many binaries to undergo chemically homogeneous evolution, and alternative, more stringent models should be considered.


\section*{Acknowledgements}

We thank Ilya Mandel, Alejandro Vigna-G\'{o}mez and Tom Callister for useful comments and discussions.
We thank the referee for a careful reading of the manuscript.
The authors are supported by the Australian Research Council (ARC) Centre of Excellence for Gravitational Wave Discovery (OzGrav), through project number CE170100004. 
SS is supported by the ARC Discovery Early Career Research Award DE220100241. T. A. C. receives support from the Australian Government Research Training Program.
This work was performed on the OzSTAR national facility at Swinburne University of Technology. 
The OzSTAR program receives funding in part from the Astronomy National Collaborative Research Infrastructure Strategy (NCRIS) allocation provided by the Australian Government.


\section*{Data availability}

This work made use of the publicly available binary population synthesis suite COMPAS \citep{COMPAS:2021methodsPaper}. COMPAS is available at \url{https://github.com/TeamCOMPAS/COMPAS}.
Results produced for this paper are available upon request.



\bibliographystyle{mnras}
\bibliography{bib} 


\bsp	
\label{lastpage}
\end{document}